\documentclass[twocolumn,showpacs,nofootinbib]{revtex4}
\usepackage[dvips]{graphics}
\usepackage{graphicx}
\usepackage{dcolumn}
\usepackage{bm}
\usepackage{amssymb,amsmath}
\usepackage{epsfig}
\usepackage{color}
\usepackage{amsfonts}
\usepackage{amscd}
\usepackage{dsfont}

\def\hatn{\mathbf{\hat n}}

\newcommand{\beq}{\begin{equation}}
\newcommand{\eeq}{\end{equation}}
\newcommand{\beqa}{\begin{eqnarray}}
\newcommand{\eeqa}{\end{eqnarray}}

\begin{document}

\title{De-Rotation of the Cosmic Microwave Background
Polarization: Full-Sky Formalism}

\author{Vera Gluscevic$^1$, Marc Kamionkowski$^1$, and Asantha Cooray$^2$}
\affiliation{$^1$ California Institute of Technology, Mail Code 350-17,
Pasadena, CA 91125\\
$^2$ Department of Physics and Astronomy, University of
California, Irvine, CA 92697}

\date{\today}

\begin{abstract}
Mechanisms  have been proposed that might rotate the
linear polarization of the cosmic microwave background (CMB) as
it propagates from the surface
of last scatter.  In the simplest scenario, the rotation will be
uniform across the sky, but the rotation angle may also vary
across the sky.
We develop in detail the complete set of full-sky quadratic estimators
for the rotation of the CMB
polarization that can be constructed from the CMB temperature
and polarization.  We derive the variance with which these estimators
can be measured and show that these variances reduce to the simpler
flat-sky expressions in the appropriate limit.  We evaluate the
variances numerically.  While the flat-sky formalism may be
suitable if the rotation angle arises as a realization of a
random field, the full-sky formalism will be required to search
for rotations that vary slowly across the sky as well as
for models in which the angular power spectrum for the rotation
angle peaks at large angles.
\end{abstract}


\maketitle
\section{Introduction}

Great strides have been made during the past decade in obtaining
precise maps of the cosmic microwave background (CMB)
temperature and polarization
\cite{deBernardis:2000gy}.  Yet there are still
increasingly precise data to come \cite{Bock:2006yf}.
Inflation \cite{Guth:1980zm}, which has passed a number
of tests so far, will be tested further
\cite{Kamionkowski:1999qc}, and we may begin to
understand the physics responsible for inflation.  Moreover,
there will be a number of different types of departures from the
standard scenario that can be detected with new measurements.

The CMB polarization pattern can be decomposed into a gradient
component (E modes) and a curl component (B modes)
\cite{Kamionkowski:1996ks,Zaldarriaga:1996}.
A given inflationary model makes specific predictions for the
power spectra of these components.  For example, if the model
predicts no primordial gravitational waves, then there is no B
mode at the CMB surface of last scatter.  If there is a
primordial gravitational-wave background with a sufficiently
large amplitude, there will be B modes at the surface of last
scatter at a detectable level \cite{Kamionkowski:1997av}, and
with a characteristic power spectrum.

Since the post-recombination Universe is transparent to the CMB, it is
natural to assume that the polarization pattern that we see is
the polarization pattern at the surface of last scatter.
However, exotic mechanisms may rotate the
linear polarization as it propagates through the Universe.  For
example, if cosmic acceleration is due to a quintessence field
$\phi$, and if that field couples to the pseudoscalar $\tilde F
F$ of electromagnetism, then the time variation of $\phi$ leads
to a rotation of the linear polarization by an angle $\alpha$
\cite{Carroll:1998zi}.  If $\phi$ is
spatially homogeneous, then the rotation angle $\alpha$ is
uniform across the sky.  In this case, there is a characteristic
parity-violating EB correlation induced in the CMB
polarization \cite{Lue:1998mq}.  Null measurements of this
effect constrain $\alpha$ to be smaller than a few degrees
\cite{Feng:2006dp,Komatsu}.

It may well be, however, that $\phi$ has spatial
variations, in which case the rotation angle $\alpha(\hatn)$
will vary with position $\hatn$ on the sky
\cite{Pospelov:2008gg}. There may also be mechanisms involving
coupling of photons to dark matter that cause a
spatially-varying $\alpha(\hatn)$ \cite{Gardner:2006za}. In either case, the EB power
spectrum would include the dependence on $\alpha(\hatn)$, which would differentiate this scenario from the uniform-rotation case. For example, the EB power spectrum even vanishes
\cite{Pospelov:2008gg} if $\alpha$ has a dipolar variation across the sky, in which case the positive EB correlation on one half of the sky is
cancelled by a negative EB correlation of equal magnitude on
the other half.

Ref.~\cite{Kamionkowski:2008fp} showed that the
rotation angle $\alpha(\hatn)$ can be reconstructed, as a
function of position on the sky, from the {\it measured}
polarization map.  If primordial perturbations are assumed to be
Gaussian (as predicted by inflation), then the rotation
introduces a characteristic non-Gaussian signal in the
polarization map.  Measurement of this non-Gaussian signal then
provides the rotation angle.  More specifically, if primordial
perturbations are Gaussian, then the spherical-harmonic
coefficients $E_{lm}$ and $B_{lm}$ of the
polarization map are statistically independent for different $l$
and $m$.  However, the rotation introduces off-diagonal
correlations, i.e. correlations between different $lm$ pairs.

Ref.~\cite{Yadav:2009eb} re-visited this proposal.  The authors worked
out the off-diagonal correlations that would be induced in the
flat-sky limit, and evaluated numerically how well the
small-angle analogues of the spherical-harmonic coefficients
$\alpha_{LM}$ of the rotation
angle could be measured with future satellite experiments like
Planck \cite{Planck} and CMBPol \cite{Bock:2008ww}.

In this paper, we work out in more detail the
full-sky formalism of Ref.~\cite{Kamionkowski:2008fp}.  We write
down the minimum-variance estimator $\hat \alpha_{LM}$ for the
rotation-angle spherical-harmonic coefficients that can be
obtained from a full-sky CMB temperature/polarization map.  We
derive expressions for the variance with which
the $\alpha_{LM}$ can be determined and show that they recover, in the large-$L$
limit, the correct flat-sky expressions.  While a flat-sky
analysis may be suitable if the rotation-angle power spectrum
peaks at small scales, the full-sky formalism will be required to
maximize the sensitivity in models, such as that in
Ref.~\cite{Pospelov:2008gg}, where the signal-to-noise peaks at
low $L$ (see, e.g., Fig.~1 in Ref.~\cite{Yadav:2009eb}).
There has also been growing attention recently to the possibility
that there may be variations in fundamental fields over distance
scales comparable to, or larger than, the horizon (perhaps remnants
of the pre-inflationary Universe) \cite{adrienne}.
Observationally, these entail
searches for departures from homogeneity/isotropy or
departures from statistical homogeneity/isotropy in the CMB or
large-scale structure \cite{Pullen:2007tu,SI}.  The full-sky
formalism we present here
can be used to search for the low-$L$ (e.g., $L=1,2,3,\cdots$)
moments of $\alpha(\hatn)$ that may arise if $\phi$ has
long-wavelength fluctuations, in addition to, or instead of, the
higher-$L$ modes that can also be probed with a survey of a
smaller region of the sky\footnote{The full-sky formalism is
exact; note, however, that it is not  computationally more
demanding than the flat-sky calculations.}.

The plan of the paper is as
follows. In \S II, we derive the temperature/polarization
correlations induced by rotation.  In \S III, we construct the
complete set of quadratic estimators for the rotation of the CMB
polarization that can be constructed from the CMB temperature and
polarization maps.  We also determine the variances with which
these estimators can be measured with a full-sky CMB map.  \S IV
derives the flat-sky limit of our variance expressions and
compares them with previous work.  \S V presents numerical
results for the variances.  Concluding remarks are provided in
\S VI.  Appendix A provides some useful formulas.  Appendix B
shows how the effects of weak lensing and rotation can be
distinguished geometrically.  Appendix C shows that our variance
expressions retrieve successfully the expressions for a uniform
rotation angle.  And Appendix D illustrates the relative
contribution of different $ll'$ pairs to the rotation-angle
estimator.


\section{Correlations Induced by Rotation} 

In this Section, we derive the temperature/polarization
correlations that are induced by a post-recombination rotation
of the polarization.  


\subsection{Induced Modes}  

We begin by recalling that the Stokes parameters $Q(\hatn)$ and
$U(\hatn)$, as a function of position $\hatn$ on the sky, are
components of a symmetric trace-free $2\times2$
tensor,\footnote{Note that the definition of $P_{ab}$ here
differs from that in Refs.~\cite{Kamionkowski:1996ks} so that
the normalization of the power spectra agree with those of
Refs.~\cite{Zaldarriaga:1996}.}
\begin{equation}
     P_{ab}(\hat{n})= \frac{1}{\sqrt{2}}
     \left(
     \begin{array}{cc}
     Q(\hat{n}) & -U(\hat{n})\sin\theta\\
     -U(\hat{n})\sin\theta & -Q(\hat{n})\sin^2\theta\\
\end{array}
\right).
\end{equation}
This tensor can be expanded in terms of tensor spherical harmonics
$Y^E_{(lm)ab}(\hatn)$ and $Y^B_{(lm)ab}(\hatn)$ in the usual
fashion \cite{Kamionkowski:1996ks,Zaldarriaga:1996},
\begin{equation}
     P_{ab} (\hat n) = \sum\limits_{l = 2}^\infty
     {\sum\limits_{m =  - l}^l {[E_{lm} Y^E _{(lm)ab} (\hat n) +
     B_{lm} Y^B _{(lm)ab}(\hat n)]} },
\label{PolarizationTensorExpansion}
\end{equation}
where $E_{lm}$ are the E-mode tensor-spherical-harmonic
coefficients, and $B_{lm}$ are the B-mode
tensor-spherical-harmonic coefficients. 

We now suppose that the polarization at the surface of last
scatter is a pure E mode; we comment on the validity of this
assumption below.  Then, a small rotation $\alpha (\hat
n)$ induces a change to the polarization,
\begin{equation}
  \delta P_{ab} (\hat n) = 2\alpha (\hat n)P_{ab}^r (\hat n) ,
\label{PolarizationChangeBySmallRotation}
\end{equation}
where \cite{Kamionkowski:2008fp}
\begin{equation}
     P_{ab}^r (\hat n) = \sum\limits_{l = 2}^\infty
     {\sum\limits_{m =  - l}^l {E_{lm} Y_{(lm)ab}^B } }.
\label{PolarizationERotatedBy45}
\end{equation}
A pure E mode thus gets rotated into a B mode.  Note that
Eq.~(\ref{PolarizationERotatedBy45}) is valid only in the limit
of small rotations, $\alpha(\hatn) \ll1$.  Given that, the B mode
will be small compared to the E mode, which satisfies existing empirical constraints.

Rotation is a scalar field on the sky and can thus be expanded in
terms of spherical harmonics,
\begin{equation}
  \alpha (\hat n) = \sum\limits_{LM} {\alpha _{LM} Y_{LM} (\hat n)}  ,
\label{RotationFieldExpansion}
\end{equation}
where $\alpha _{LM}$ are the coefficients of the expansion.

Using Eqs.~(\ref{PolarizationChangeBySmallRotation}),
(\ref{PolarizationERotatedBy45}), and
(\ref{RotationFieldExpansion}), the B mode induced by a small
rotation angle from a pure E mode can be expressed as
\begin{equation}
\begin{gathered}
     \delta B_{lm}  = \int {d\hat n\delta P_{ab} (\hat
     n)Y^{B*,ab} _{(lm)} } (\hat n) \hfill \\  
     =2 \sum\limits_{LM} {\sum\limits_{l_2 m_2 } {\alpha _{LM}
     E_{l_2 m_2 } \int {d\hat n} } } Y^{B*,ab} _{(lm)} Y_{(LM)}
     Y^B _{(l_2 m_2 )ab} .
\end{gathered}
\label{InducedBMode_FromPureE_General}
\end{equation}
Similarly, the induced E mode is
\begin{equation}
\begin{gathered}
     \delta E_{lm}  = \int {d\hat n\delta P_{ab} (\hat
     n)Y^{E*,ab} _{(lm)} } (\hat n) \hfill \\ 
     = 2\sum\limits_{LM} {\sum\limits_{l_2 m_2 } {\alpha _{LM} E_{l_2
     m_2 } \int {d\hat n} } } Y^{E*,ab} _{(lm)} Y_{(LM)} Y^B _{(l_2
     m_2 )ab} . 
\end{gathered}
\label{InducedEMode_FromPureE_General}
\end{equation}

We further develop the last two expressions using
\cite{Kamionkowski:1996ks}     
\begin{equation}
     Y_{(lm)ab}^B = \frac{{{N_l}}}
     {2}({Y_{(lm):ac}}{\varepsilon ^c}_b + {Y_{(lm):bc}}{\varepsilon ^c}_a) ,
\label{YB_DerivativesOfYs}
\end{equation}
\begin{equation}
Y_{(lm)ab}^E = {N_l}({Y_{(lm):ab}} - \frac{1}
{2}{g_{ab}}{Y_{(lm):c}}^c),
\label{YE_DerivativesOfYs}
\end{equation}
where $g$ and $\varepsilon$ are, respectively, the metric tensor
and the Levi-Civita tensor on a unit 2-sphere (see Appendix A);
a colon denotes a covariant derivative on the 2-sphere; and
$N_l$ is given by 
\begin{equation}
N_l  \equiv \sqrt {\frac{{2(l - 2)!}}
{{(l + 2)!}}} .
\label{NL}
\end{equation}
From Ref.~\cite{Hu:2000ee}, we can express the double
derivatives in terms of spin-2 spherical harmonics
${}_{\pm2}Y_{(lm)}(\hatn)$ as
\begin{equation}
\begin{gathered}
     {Y_{(lm):ab}} =  - \frac{{l(l + 1)}}
     {2}{Y_{(lm)}}{g_{ab}} + \frac{1}
     {2}\sqrt {\frac{{(l + 2)!}}
     {{(l - 2)!}}} \\
     \times [_2 Y_{(lm)} (m_+\otimes m_+)  + _{ - 2} Y_{(lm)} (m_-
     \otimes m_-)  ]_{ab},
\end{gathered}
\label{DerivativesOfYs_mVectors}
\end{equation}
where
\begin{equation}
m_ \pm   = \frac{1}
{{\sqrt 2 }}(\hat e_\theta   \mp i\hat e_\phi  ).
\label{mVectors}
\end{equation}

Combining Eqs.~(\ref{DerivativesOfYs_mVectors}) and (\ref{mVectors})
with Eqs.~(\ref{YB_DerivativesOfYs}), (\ref{YE_DerivativesOfYs}),
and (\ref{NL}), we can express the tensor spherical harmonics
${Y^{E,B}_{(lm)}}$ in terms of spin-2 spherical
harmonics as\footnote{We suppress the $(lm)$ indices for $Y$ in
this formula.}
\begin{equation}
\begin{gathered}
  Y^B _{(lm)}  =\hfill \\
\left( {\begin{array}{*{20}c}
   {\frac{i\sqrt 2}
{4}(_{+2}Y_{}  - _{-2}Y_{} )} & {\frac{\sqrt 2}
{4}\sin (\theta )(_{-2}Y_{}  + _{+2}Y_{} )}  \\
   {\frac{\sqrt 2}
{4}\sin (\theta )(_{-2}Y_{}  + _{+2}Y_{} )} & {\frac{i\sqrt 2}
{4}\sin ^2 (\theta )(_{-2}Y_{}  - _{+2}Y_{} )}  \\
 \end{array} } \right) ,
\end{gathered}\label{YBMatrix_Yspin2}
\end{equation}
\begin{equation}
\begin{gathered}
 Y^E _{(lm)}  =\hfill \\
\left( {\begin{array}{*{20}c}
   {\frac{\sqrt 2}
{4}(_{+2}Y_{}  + _{-2}Y_{} )} & {\frac{i\sqrt 2}
{4}\sin (\theta )(_{-2}Y_{}  - _{+2}Y_{} )}  \\
   {\frac{i\sqrt 2}
{4}\sin (\theta )(_{-2}Y_{}  - _{+2}Y_{} )} & {-\frac{\sqrt 2}
{4}\sin ^2 (\theta )(_{-2}Y_{}  + _{+2}Y_{} )}  \\
 \end{array} } \right).
\end{gathered}
\label{YEMatrix_Yspin2}
\end{equation}
Using this result, we obtain
\begin{equation}
\begin{gathered}
Y^{B*, ab} _{(lm)} Y^B _{(l_2 m_2 )ab}  \hfill \\
= \frac{1}
{2}(_{ - 2} Y_{(lm)} ^*  \times _{ - 2} Y_{(l_2 m_2 )}  + _{+2}Y_{(lm)} ^*  \times _{+2}Y_{(l_2 m_2 )} ),
\end{gathered}
\label{YBstarYB_Yspin2}
\end{equation}
and
\begin{equation}
\begin{gathered}
  Y^{E*, ab} _{(lm)} Y^B _{(l_2 m_2 )ab}  \hfill \\
   = \frac{i}
{2}(  _{ + 2} Y_{(lm)} ^*  \times _{ + 2} Y_{(l_2 m_2 )} - _{-2} Y_{(lm)} ^*  \times _{-2} Y_{(l_2 m_2 )} ) .
\end{gathered} 
\label{YEstarYB_Yspin2}
\end{equation}

The next step is to use Eqs.~(\ref{YBstarYB_Yspin2}) and
(\ref{YEstarYB_Yspin2}) in order to rewrite the integrals in
Eqs.~(\ref{InducedBMode_FromPureE_General}) and
(\ref{InducedEMode_FromPureE_General}) in terms of Wigner 3j
symbols. We use Ref.~\cite{Hu:2000ee}, noting that the
spin-spherical harmonics of zero spin are the regular spherical
harmonics, $_{0}Y_{(lm)} \equiv Y_{(lm)}$. Let us first look at
the case of the induced B mode, where we have 
\begin{equation}
\begin{gathered}
  \int {d\hat nY^{B*,ab} _{(lm)} Y_{(LM)} Y^B _{(l_2 m_2 )ab} } \hfill \\
   = \frac{1}
{2}( - 1)^m \sqrt {\frac{{(2l + 1)(2L + 1)(2l_2  + 1)}}
{{4\pi }}} \hfill \\
\left[ {\left( {\begin{array}{*{20}c}
   l & L & {l_2 }  \\
   { - 2} & 0 & 2  \\
 \end{array} } \right) + \left( {\begin{array}{*{20}c}
   l & L & {l_2 }  \\
   2 & 0 & { - 2}  \\
 \end{array} } \right)} \right] \left( {\begin{array}{*{20}c}
   l & L & {l_2 }  \\
   { - m} & M & {m_2 }  \\
 \end{array} } \right). 
\end{gathered}
\label{IntegralYBstarYBY} 
\end{equation}

We now define\footnote{Note that the definitions of
$\xi^{LM}_{lml'm'}$ and $H^L_{ll'}$ differ from those in
Ref.~\cite{Kamionkowski:2008fp}.  We define these quantities in
this way to avoid division by zero.}
\begin{equation}
\begin{gathered}
\xi _{lm{l_2}{m_2}}^{LM} \equiv {( - 1)^m}\sqrt {\frac{{(2l + 1)(2L + 1)(2{l_2} + 1)}}
{{4\pi }}} {\text{ }}\hfill \\
\times \left( {\begin{array}{*{20}{c}}
   l & L & {{l_2}}  \\
   { - m} & M & {{m_2}}  \\
 \end{array} } \right),
\end{gathered}
\label{KsiDefinition_Wigner3j}
\end{equation}
and
\begin{equation}
H_{ll_2 }^L  \equiv \left( {\begin{array}{*{20}c}
   l & L & {l_2 }  \\
   { 2} & 0 & {-2}  \\

 \end{array} } \right) .
\label{HDefinition_Wigner3j}
\end{equation}
Also, note that, due to the properties of the Wigner 3j
symbols,\footnote{See Appendix A: changing the sign on all three
$m$'s, brings up a factor of $(-1)^{l+l_2+L}$.} the sum in
Eq.~(\ref{IntegralYBstarYBY}) vanishes, unless $l + l_2 +
L=$ even. Replacing Eqs.~(\ref{IntegralYBstarYBY}),
(\ref{KsiDefinition_Wigner3j}), and (\ref{HDefinition_Wigner3j}) in
Eq.~(\ref{InducedBMode_FromPureE_General}), we come to a
relatively simple expression for the rotation-induced B mode,
\begin{equation}
{\delta B_{lm}  =2 \sum\limits_{LM} {\sum\limits_{l_2 m_2 } {\alpha _{LM} E_{l_2 m_2 } } } \xi _{lml_2 m_2 }^{LM} H_{ll_2 }^L },
\label{InducedBMode_FromPureE}
\end{equation} 
where the only non-zero terms in the sum are those that satisfy $l+l_2+L =$ even.

Similarly, for the case of the rotation-induced E mode, again using the
properties of the Wigner 3j symbols,\footnote{See Appendix A: if
the sum of $m$'s does not vanish, the value of the symbol is
zero.} we have
\begin{equation}
\int {d\hat nY^{E*,ab} _{(lm)} Y_{(LM)} Y^B _{(l_2 m_2 )ab} }  = i\xi _{lml_2 m_2 }^{LM} H_{ll_2 }^L ,
\label{IntegralYEstarYBY}
\end{equation}
and then replacing this in Eq.(~\ref{InducedEMode_FromPureE_General}), we get
\begin{equation}
\delta E_{lm}  = 2i\sum\limits_{LM} {\sum\limits_{l_2 m_2 } {\alpha _{LM} E_{l_2 m_2 } } } \xi _{lml_2 m_2 }^{LM} H_{ll_2 }^L .
\label{InducedEMode_FromPureE}
\end{equation}
Note that in Eqs.~(\ref{IntegralYEstarYBY}) and
(\ref{InducedEMode_FromPureE}) the only non-zero terms are
those that satisfy $l+l_2+L=$ odd.


\subsection{Induced Correlations}

Using the induced modes, derived in the previous subsection, we
derive four correlations that are modified/induced by the rotation: EB,
EE, TB, and TE (there is also a BB correlation, but it is higher
order in $\alpha$, and thus neglected). 
Every mode that we detect will contain the sum of the primordial (at the surface of last scatter) and the rotation-induced component,
\begin{equation}
\begin{gathered}
  {E_{lm}} = {E_{lm,0}} + \delta {E_{lm}}, \hfill \\
  {B_{lm}} = \delta {B_{lm}}.
\end{gathered}
\label{DetectedModes}
\end{equation}
We use the definitions of the EE and TE power spectra,
\begin{equation}
\begin{gathered}
  \left\langle {{E_{lm,0}}E_{l'm',0}^*} \right\rangle  = C_l^{EE}\delta _{ll'}\delta _{mm'} ,\hfill \\
  \left\langle {{E_{lm,0}}T_{l'm',0}^*} \right\rangle  = C_l^{TE}\delta _{ll'}\delta _{mm'} .\hfill \\ 
\end{gathered} 
\label{PowerSpectra_EEandET}
\end{equation}

Finally, we obtain the expressions for the four
correlators, to first order in $\alpha$. Using
Eqs.~(\ref{InducedBMode_FromPureE}) and
(\ref{InducedEMode_FromPureE})--(\ref{PowerSpectra_EEandET}), we
obtain
\begin{equation}
\begin{gathered}
\left\langle {B_{lm} E_{l'm'}^* } \right\rangle  = 2\sum\limits_{LM} { {\alpha _{LM} C_{l'}^{EE} \xi _{lml'm'}^{LM} H_{ll'}^L } } \hfill \\ 
= \frac{1}
{{\sqrt \pi  }}\alpha _{00} C_{l'}^{EE} \delta _{ll'} \delta _{mm'} \hfill \\
 + 2\sum\limits_{L \geqslant 1} {\sum\limits_{M =  - L}^L {\alpha _{LM} C_{l'}^{EE} \xi _{lml'm'}^{LM} H_{ll'}^L } } ,
\end{gathered}
\label{InducedCorrelator_EB}
\end{equation}
and the rest are
\begin{equation}
\begin{gathered}
\left\langle {{E_{lm}}E_{l'm'}^*} \right\rangle  = C_l^{EE}\delta _{ll'}\delta _{mm'} \hfill \\
 + 2i\sum\limits_{LM} {(C_{l'}^{EE} - C_l^{EE}){\alpha _{LM}}\xi _{lml'm'}^{LM}H_{ll'}^L} ,
\end{gathered}
\label{InducedCorrelator_EE}
\end{equation}
\begin{equation}
\left\langle {B_{lm} T_{l'm'}^* } \right\rangle  = 2\sum\limits_{LM} {{\alpha _{LM} C_{l'}^{TE} \xi _{lml'm'}^{LM} H_{ll'}^L  } },
\label{InducedCorrelator_BT}
\end{equation}
\begin{equation}
\begin{gathered}
\left\langle {{E_{lm}}T_{l'm'}^*} \right\rangle  = C_l^{TE}\delta _{ll'}\delta _{mm'}\hfill \\
+ 2i\sum\limits_{LM} {C_{l'}^{TE}{\alpha _{LM}}\xi _{lml'm'}^{LM}H_{ll'}^L}.
\end{gathered}
\label{InducedCorrelator_TE}
\end{equation}

Note that for EB and TB, the sum is taken over the terms that satisfy $l+l'+L=$ even and in EE and TE over $l+l'+L=$ odd.


\section{Estimators for the Rotation-angle Spherical-Harmonic
Coefficients}

If we assume that the primordial CMB
temperature/polarization pattern is a realization of a
statistically isotropic Gaussian random field, then the
spherical-harmonic coefficients ($T_{lm}$, $B_{lm}$, or
$E_{lm}$) for the {\it primordial} field are all uncorrelated.
As Eqs.~(\ref{InducedCorrelator_EB})--(\ref{InducedCorrelator_TE})
show however, rotation induces {\it off-diagonal} correlations;
i.e., correlations between spherical-harmonic coefficients of
different $lm$ and $l'm'$.

\begin{table}[tbp]
\begin{center}
\begin{tabular}{c|c}
A  & $Z_{ll'}^A$ \\
\hline
BE & $C_{l'}^{EE}$ \\
EB & $C_{l}^{EE}$ \\
EE & $-i (C_{l'}^{EE}-C_l^{EE})$ \\
BT & $C_{l'}^{TE}$ \\
TB & $C_{l}^{TE}$ \\
ET & $-i C_{l'}^{TE}$ \\
TE & $ -i C_{l}^{TE}$ \\
\end{tabular}
\caption{The quantities $Z_{ll'}^A$, defined in
     Eq.~(\protect\ref{DDefinition_inEB}), for the various modes A.}
\end{center}
\end{table}

While the correlations of specific $lm$-$l'm'$ pairs depend on the
azimuthal quantum numbers $m$ and $m'$, they can be
parametrized in terms of the rotational invariants (i.e.,
independent of $m$),
\begin{equation}
     D_{ll'}^{LM,A} \equiv 2{\alpha _{LM}} Z_{ll'}^{A} H_{ll'}^L,
\label{DDefinition_inEB}
\end{equation}
where the quantities $Z_{ll'}^A$ are given in Table 1; they are
obtained from
Eqs.~(\ref{InducedCorrelator_EB})--(\ref{InducedCorrelator_TE}).
The
EB correlator, for a given $ll'$ pair with $l\neq l'$, is
different than the BE correlator.  We thus consider {\it both}
BE and EB (and similarly for TE/ET and TB/BT) and then restrict
our sums to $l'\geq l$ to avoid double-counting.  We thus have
for $l> l'$, $A=\{BE,EB,TB,BT,TE,ET,EE\}$, while for $l=l'$, we
have $A=\{BE,TE,TB,EE\}$.  With these shorthands, the part of
the $XX'$ correlators (for $\{X,X'\} = \{T,B,E\}$) can be written as
\begin{equation}
     \left\langle {{X_{lm}}(X'_{l'm'})^*} \right\rangle  =
     \sum\limits_{LM} D_{ll'}^{LM,XX'} \xi _{lml'm'}^{LM}.
\label{EB_ksiD}
\end{equation}

We now suppose that we have spherical-harmonic coefficients
$T_{lm}^{\text{map}}$, $E_{lm}^{\text{map}}$,
$B_{lm}^{\text{map}}$, obtained from a CMB
temperature/polarization map.  These receive contributions from
the true signal on the sky, reduced by the $l$-space window
function $W_l = \exp ( - {l^2}\sigma _b^2/2)$, where
$\sigma_b=\theta_{\mathrm{fwhm}}/\sqrt{8\ln 2} =
0.00741\,(\theta_{\mathrm{fwhm}}/1^\circ)$ for a Gaussian beam of
width $\theta_{\text{fwhm}}$, and a contribution from
detector noise.  The predictions for the rotational invariants
for the {\it map} are $D_{ll'}^{LM,XX',\text{map}} =
D_{ll'}^{LM,XX'} W_l W_{l'}$.  Following
Refs.~\cite{Pullen:2007tu,Kamionkowski:2008fp,Hajian:2005jh},
the minimum-variance estimator for each
$D_{ll'}^{LM,XX',\text{map}}$ is\footnote{Note that the
definition of $G^L_{ll'}$ differs from that in
Refs.~\cite{Pullen:2007tu,Kamionkowski:2008fp}.}
\begin{equation}
     {\hat D_{ll'}^{LM,XX',\text{map}} = {(G_{ll'}^L)^{ -
     1}}\sum\limits_{mm'} {
     {X_{lm}^{\text{map}} X_{l'm'}^{\prime,\text{map},*}} } \xi
     _{lml'm'}^{LM}},
\label{DEstimator_inEB}
\end{equation}
where we have used (see Appendix A),
\begin{equation}
     \sum\limits_{mm'} {} {(\xi _{lml'm'}^{LM})^2} = G_{ll'}^L
     \equiv  \frac{{(2l + 1)(2l' + 1)}} {{4\pi}}.
\label{Ksi_G}
\end{equation}
Recall also that for EB and TB, only the terms in
Eq.~(\ref{DEstimator_inEB}) that satisfy
$l+l'+L=$even are non-vanishing, while for EE and TE only
$l+l'+L=$odd terms are non-vanishing.

The variances with which each $\hat
D_{ll'}^{LM,XX',\text{map}}$ can be measured can also be
calculated.  Moreover, measurements of different $\hat
D_{ll'}^{LM,XX',\text{map}}$ will be correlated, for the same
$ll'$.  We write the covariances between the $\hat
D_{ll'}^{LM,XX',\text{map}}$ in terms of the quantities,
\begin{eqnarray}
     \mathcal{C}_{AA'}^{ll'} &\equiv&  {G_{ll'}^L} \left( \left\langle
     \hat D_{ll'}^{LM,A,\text{map}} \hat
     D_{ll'}^{LM,A',\text{map}} \right\rangle \right. \nonumber \\
     & & \left. - \left\langle \hat D_{ll'}^{LM,A,\text{map}}
     \right\rangle \left\langle \hat D_{ll'}^{LM,A'\text{map}}
     \right\rangle \right).
\label{CurlyC_GD}
\end{eqnarray}
In principle, $\mathcal{C}_{AA'}^{ll'}$ is a $7\times7$ matrix
(in the $AA'$ space) for $l\neq l'$ and $4\times4$ for $l=l'$.
However, the matrix is sparsely populated in the $AA'$ space; it can be written in
block-diagonal form, since the EB and TB correlators are
non-vanishing only for $l+l'+L=$even while those for TE and EE
are non-vanishing only for $l+l'+L=$odd.  We present
explicit expressions for the relevant entries of
$\mathcal{C}_{AA'}^{ll'}$ below.

\subsection{Minimum-Variance estimators for the rotation-angle
coefficients}

We now write down the minimum-variance quadratic estimator that
can be constructed for the rotation-angle coefficients
$\alpha_{LM}$ from the measured temperature/polarization map.

To begin, we recall that each $\hat D_{ll'}^{LM,A,\text{map}}$
(i.e., each $ll'$ and $AA'$, for a given $LM$) provides a
measurement of $\alpha_{LM}$, through $(\hat\alpha_{LM})_{ll'}^A = \hat
D_{ll'}^{LM,A,\text{map}}/F_{ll'}^{L,A}$, in terms of the quantities
\begin{equation}
     F_{ll'}^{L,A} \equiv 2 Z_{ll'}^A H_{ll'}^{L} W_l W_{l'}.
\label{FDefinition_inEB}
\end{equation}
The variances and covariances of the $(\hat \alpha_{LM})_{ll'}^A$ are given simply in terms of those for
$D_{ll}^{LM,A,\text{map}}$ scaled by the appropriate factors of
$F_{ll'}^{L,A}$.

The minimum-variance estimator $\hat \alpha_{LM}$ is then
obtained by summing all of the individual estimators, for a
given $LM$, with inverse-variance weighting (and taking into
account also the covariances).  The minimum-variance
estimator is thus
\begin{equation}
     \hat \alpha_{LM} =  \sigma_{\alpha_{LM}}^2  \sum_{l' \geq l} G^L_{ll'}
     \sum_{AA'} F_{ll'}^{L,A'} \hat D_{ll'}^{LM,A,\text{map}}
     \left [ (\mathcal{C}^{ll'})^{-1} \right]_{AA'},
\label{formalestimator}
\end{equation}
and it has variance $\sigma_{\alpha_{LM}}^{2}$ given by
\begin{equation}
     \sigma_{\alpha_{LM}}^{-2}=
     \sum_{\l' \geq l} G^L_{ll'} \sum_{AA'} F_{ll'}^{L,A} F_{ll'}^{L,A'}
      \left [ (\mathcal{C}^{ll'})^{-1}
     \right]_{AA'}.
\label{formalvariance}
\end{equation}
In these expressions, the $AA'$ sums are over
$\{EB,BE,TB,BT,EE,TE,ET\}$ for $l\neq l'$ and  $\{EB,TB,EE,TE\}$
for $l=l'$, and the matrix inversion is in the $AA'$ space.


\subsection{Variance for $D$}

We assume that the noise is isotropic and that it is not correlated with the data; that
the temperature and polarization noises are  not correlated; and
that the temperature/polarization noises between different
pixels are not correlated.  If so, then the power spectra for
the map are $C_{l}^{A,\text{map}} = {C_{l}^{A}}{W_l^2} +
C_{l}^{A, \text{noise}}$, where here A$= \{\text{TT, EE, BB}\}$,
as described in Ref.~\cite{Kamionkowski:2008fp}. The TT, EE, and
BB noise power spectra are
\begin{equation}
\begin{gathered}
     C_l^{TT,\text{noise}} = (4\pi /{N_{\text{pix}}}){\sigma
     _T}^2 ,\hfill \\
     C_l^{BB,\text{noise}} = C_l^{EE,\text{noise}} = (4\pi
     /{N_{\text{pix}}}){\sigma _P}^2,
\end{gathered}
\label{Noise_WMAP}
\end{equation}
where $\sigma_T^2$ and $\sigma_P^2$ are the temperature and
noise variances in each pixel, and $N_{\text{pix}}$ is the number
of pixels in the map.  The instrumental-noise contributions to
the cross-power spectra are zero:
\begin{equation}
     C_l^{EB,\text{noise}} = C_l^{TB,\text{noise}} =
     C_l^{TE,\text{noise}} = 0.
\label{Noise_EB_TB_TE}
\end{equation} 

Now we can calculate the desired variances for $D$, using
Eqs.~(\ref{DEstimator_inEB}) and the power spectra $C_l^A$.
For EB, for $l\neq l'$, we have
\begin{equation}
\begin{gathered}
  \mathcal{C}_{BE,BE}^{ll'} = C_l^{BB,\mathrm{map}}C_{l'}^{EE,\mathrm{map}}, \hfill \\
  \mathcal{C}_{EB,EB}^{ll'} = C_{l'}^{BB,\mathrm{map}}C_l^{EE,\mathrm{map}}, \hfill \\
  \mathcal{C}_{EB,BE}^{ll'} = \mathcal{C} _{BE,EB}^{ll'} = 0, \hfill \\ 
\end{gathered} 
\label{CurlyC_inEB_lnotlp}
\end{equation}
and for $l=l'$,
\begin{equation}
\mathcal{C}_{BE,BE}^{ll}
=C_l^{BB,\mathrm{map}}C_{l}^{EE,\mathrm{map}}.
\end{equation}
The covariances for BT and TB are the same, with the
replacements E$\to$T.  There are also covariances between the TB
and EB estimators.  For $l=l'$, they are
\begin{equation}
  \mathcal{C}_{BE,BT}^{ll} =
  C_l^{BB,\mathrm{map}}C_{l}^{TE,\mathrm{map}}.
\end{equation}
For $l\neq l'$,
\begin{equation}
\begin{gathered}
  \mathcal{C}_{BE,BT}^{ll'} = C_l^{BB,\mathrm{map}}C_{l'}^{TE,\mathrm{map}}, \hfill \\
  \mathcal{C}_{EB,TB}^{ll'} = C_{l'}^{BB,\mathrm{map}}C_l^{TE,\mathrm{map}}, \hfill \\
  \mathcal{C}_{BE,TB}^{ll'} = \mathcal{C}_{EB,BT}^{ll'} = 0. \hfill \\ 
\end{gathered} 
\label{CurlyC_inTBEB_lnotlp}
\end{equation}
For EE,
\begin{equation}
     \mathcal{C} _{EE,EE}^{ll'} =
     (1+\delta_{ll'}) C_l^{EE,\mathrm{map}}C_{l'}^{EE,\mathrm{map}}.
\label{CurlyC_inEE}
\end{equation}
For the TE case, for $l\neq l'$, we have
\begin{equation}
\begin{gathered}
  \mathcal{C} _{TE,TE}^{ll'} = C_l^{TT,\mathrm{map}}C_{l'}^{EE,\mathrm{map}}, \hfill \\
  \mathcal{C} _{ET,ET}^{ll'} =
  C_{l'}^{TT,\mathrm{map}}C_l^{EE,\mathrm{map}}, \hfill \\
  \mathcal{C} _{TE,ET}^{ll'} = \mathcal{C} _{ET,TE}^{ll'} =
  C_l^{TE,\mathrm{map}}C_{l'}^{TE,\mathrm{map}},\hfill \\ 
\end{gathered}
\label{CurlyC_inTE_lnotlp}
\end{equation}
and for $l=l'$,
\begin{equation}
  \mathcal{C} _{ET,ET}^{ll} =
  C_{l}^{TT,\mathrm{map}}C_l^{EE,\mathrm{map}} 
  \left( C_l^{TE,\text{map}} \right)^2.
\end{equation}

There are also TE-EE covariances.  However, since TE and EE are
almost always weaker probes of the rotation, we do not include
these additional expressions here.


\subsection{Quadratic Estimators for BE (or TE) only}

As an example (and for clarity), we can write down the expressions
for the estimator and noise in the case where we use only
information from the BE correlator to determine $\alpha_{LM}$.
In this case, the estimator is
\begin{eqnarray}
     \hat\alpha_{LM} &=& \sigma_{\alpha_{LM}}^2 \sum_{l'\geq l}
     (1+\delta_{ll'})^{-1} G^L_{ll'} \nonumber \\
     & & \times \left[ \frac{F_{ll'}^{L,BE}
     D_{ll'}^{LM,BE,\text{map}}}{ C_l^{BB,\text{map}}
     C_{l'}^{EE,\text{map}}} + ( \text{B} \leftrightarrow
     \text{E} ) \right],
\label{BEestimator}
\end{eqnarray}
and the noise is given by
\begin{eqnarray}
     {\sigma_{\alpha_{LM}}^{-2}}& =&  \sum_{l'\geq l}
     (1+\delta_{ll'})^{-1} G^L_{ll'}  \nonumber \\
     & & \times \left[ \frac{(F_{ll'}^{L,BE})^2}{ C_l^{BB,\text{map}}
     C_{l'}^{EE,\text{map}}} + ( \text{B} \leftrightarrow
     \text{E} ) \right].
\label{BEnoise}
\end{eqnarray}
The $ll'$ sums here are over values that satisfy $l+l'+L=$even.
The estimator and variance for TB are the same after the
replacement E$\to$T.


\section{Flat-sky limit and comparison with previous work}

In this Section, we derive the flat-sky limit of the variances
for the rotation on the full sky and compare our results to
those of Ref.~\cite{Yadav:2009eb}. We work out the EB case,
where $l+l'+L=$ even. The other three cases follow analogously.

From Eqs.~(\ref{BEnoise}) and (\ref{FDefinition_inEB}), the variance is 
\begin{equation}
\begin{gathered}
     {\sigma _L}^{ - 2} = 4\sum_{l' > l}
     (H_{ll'}^L)^2 G_{ll'}^L{({W_l}{W_{l'}})^2}\hfill
     \\ \times
     \left[\frac{{{{(C_{l'}^{EE})}^2}}}
     {{C_l^{BB,{\text{map}}}C_{l'}^{EE,{\text{map}}}}} +
     \frac{{{{(C_l^{EE})}^2}}}
     {{C_{l'}^{BB,{\text{map}}}C_l^{EE,{\text{map}}}}} \right].
\end{gathered}
\label{VarianceOfRotation_lnotlp_inEB}
\end{equation}
The two terms in Eq.~(\ref{VarianceOfRotation_lnotlp_inEB}) are
the same under the exchange of  $l$ and $l'$. Thus, after
renaming the indices on one of the two terms, we get the sums
over $l<l'$ and $l>l'$, which covers the whole range of
$l$'s\footnote{Note that, when we switch to integration, as
shown further on in the text, the $l=l'$ term is of measure
zero, and can be ignored.}. We are left with
\begin{equation}
     {(\sigma _L)}^{ - 2} = 4\sum\limits_{ll'} {} X_{ll'}^L
     G_{ll'}^L(H_{ll'}^L)^2,
\label{sigma_XGH_in_EB}
\end{equation}
where we have defined
\begin{equation}
     X_{ll'}^L \equiv ({W_l}{W_{l'}})^2\frac{{{{( C_{l'}^{EE})}^2}}}
     {{C_{l'}^{EE,\text{map}}C_l^{BB,\text{map}}}}.
\label{XDefinition_inEB}
\end{equation}

We now derive the limit of high multipoles.\footnote{If $L$ is
large, then the triangle inequalities and the requirement for
non-flat triangles ensures that $l$ and $l'$ are also
large.}  We start by using the approximation \cite{Hu:2000ee}, 
\begin{equation}
H_{ll'}^L \approx \cos 2{\varphi _{ll'}}\left( {\begin{array}{*{20}{c}}
   l & {l'} & L  \\
   0 & 0 & 0  \\
 \end{array} } \right),
\label{H_cos}
\end{equation}
for the $L+l+l'=$even case. 
From Eqs.~(\ref{Ksi_G}) and (\ref{H_cos}), we have
\begin{equation}G_{ll'}^L(H_{ll'}^L)^2 \xrightarrow{L,l,l' \to \infty } \frac{ll'}
{\pi }{\left( {\begin{array}{*{20}{c}}
   l & {l'} & L  \\
   0 & 0 & 0  \\

 \end{array} } \right)^2}\cos ^2 2{\varphi _{ll'}}.
\label{GsqH_HighLLimit}
\end{equation}
From the relation of the spherical harmonics and Wigner 3j
symbols (see Appendix A), this gives, for large $L$,
\begin{equation}
\begin{gathered}
{(\sigma _L)}^{ - 2}\xrightarrow[{L,l,l' \to \infty }]{{}}4\sum\limits_{ll'} {} X_{ll'}^L\sqrt {\frac{{ll'}}
{{2\pi L}}}\hfill \\
 \times\int d\hat nY_{l0}{Y_{l'0}}{Y_{L0}}\cos ^2 2{\varphi _{ll'}}.
\end{gathered}
\label{sigma_inEB_LargeLLimit}
 \end{equation}
Given that
\begin{equation}
\int\limits_0^{2\pi } {\frac{{d\varphi_l }}
{{2\pi }}{e^{im{\varphi _l}}} = {\delta _{m,0}}},
\label{DeltaDefinition}
 \end{equation}
Eq.~(\ref{sigma_inEB_LargeLLimit}) can be re-written using
\begin{equation}
\begin{gathered}
\sum\limits_{mm'M} {\sum\limits_{ll'} {\sqrt {\frac{{ll'}}
{{2\pi L}}} \int {d\hat nY_{lm}^*} {Y_{l'm'}}{Y_{LM}}{\delta _{M,0}}{\delta _{m,0}}{\delta _{m',0}}} } \hfill \\
 = \sum\limits_{mm'M} \sum\limits_{ll'} \sqrt {\frac{{ll'}}
{{2\pi L}}} \int {d\hat nY_{lm}^*} {Y_{l'm'}}{Y_{LM}} \hfill \\
\int {\frac{{d{\varphi _l}d{\varphi _{l'}}d{\varphi _L}}}
{{{{(2\pi )}^3}}}} {e^{i(M{\varphi _L} + m'{\varphi _{l'}} - m{\varphi _l})}}.
\end{gathered}
\label{tripleIntegral_withDeltas_LargeLLimit}
\end{equation}

We use relations from Ref.~\cite{Hu:2000ee} 
\begin{equation}
{e^{i\cdot \vec{ l}\cdot{\vec {n}}}} \approx \sqrt {\frac{{2\pi }}
{l}} \sum\limits_m {{i^m}{Y_{lm}}{e^{im{\varphi _l}}}} ,
\label{Delta_sumYs}
\end{equation}
\begin{equation}
\begin{gathered}
\delta (\vec L - (\vec l - \vec l')) = \int {\frac{{d\hat n}}
{{{{(2\pi )}^2}}}} {e^{i(\vec L - \vec l + \vec l') \cdot \hat n}} \hfill \\
\approx \int {\frac{{d\hat n}}
{{{{(2\pi )}^2}}}} \sum\limits_{mm'M} {Y_{lm}^*{Y_{l'm'}}{Y_{LM}}\sqrt {\frac{{{{(2\pi )}^3}}}
{{ll'L}}} }\hfill \\
 {e^{i(M{\varphi _L} + m'{\varphi _{l'}} - m{\varphi _l})}},
\end{gathered}
\label{Delta_sumIntegralYs}
\end{equation}
and replace the sum with the integral, 
\begin{equation}
\sum\limits_{ll'} {\int\, {d{\varphi _l}\, d{\varphi _{l'}}\, ll'}
}  \leftrightarrow \int\, {\int {{d^2}\vec l\, {d^2}\vec l'} }.
\label{sumIntegral_doubleIntegral}
 \end{equation}
From
Eqs.~(\ref{tripleIntegral_withDeltas_LargeLLimit})--(\ref{sumIntegral_doubleIntegral})
and Eq.~(\ref{sigma_inEB_LargeLLimit}), and after integrating
over $d\varphi _L$, we obtain the flat-sky limit for variance in
the EB case,
\begin{equation}
\begin{gathered}
     {(\sigma _L)}^{-2}\xrightarrow[{L,l,l' \to \infty }]{}
     4\int {\frac{{{d^2}l'}} 
     {{{{(2\pi )}^2}}}} {\cos ^2}2{\varphi _{l'l}} \hfill \\
     \times(W_lW_{l'})^2\frac{(
     C_{l'}^{EE})^2}{C_l^{BB,\text{map}}C_{l'}^{EE,\text{map}}}.
\end{gathered}
\label{VarianceOFRotation_FlatSkyLimit}
\end{equation}
This can be shown to agree with the results of
Ref.~\cite{Yadav:2009eb} after combining their Eqs.~(7) and (8)
and results from their Table 1.


\section{Numerical Results}

We now present numerical results for the variances of the
estimators for a position-dependent rotation,
for different instruments.  The primordial power spectra, at the
surface of last scatter, are obtained using WMAP-5 cosmological
parameters: $\Omega
_bh^2=0.02267$, $\Omega_ch^2=0.1131$, $\Lambda=0.726$,
$n_s=0.960$, $\tau=0.084$, and a power spectrum normalized to
WMAP5~\cite{Komatsu}.  

We analyze three different experiments: (i) CMBPol's (EPIC-2m) 150 GHz
channel with resolution $\theta _{\text{fwhm}} = 5'$, taking the
relevant parameters as given in Ref.~\cite{Betoule:2009pq}, a
noise-equivalent temperature NET$= 2.8\mu
\text{K}\sqrt{\text{sec}}$ and the observation time $t_{\text{obs}} =4$
yr; (ii) Planck 143 GHz channel, with $\theta _{\text{fwhm}} =
7.1'$, NET $= 31 \mu \text{K}\sqrt{\text{sec}}$ and
$t_{\text{obs}} = 1.2$ yr; (iii) WMAP, with $\theta _{\text{fwhm}} =
21'$, $\sigma _T = 30 \mu \text{K}$ and $\sigma _P = 42.6 \mu
\text{K}$ \cite{Pullen:2007tu}.
The NET parameters specified for Planck and CMBPol are related
to the temperature/polarization pixel-noise variances through
$\sigma_T^2/N_{\text{pix}}= f_{\text{sky}} (\text{NET})^2/t_{\text{obs}}$,
where $f_{\text{sky}}$ is the fractional sky coverage (here
assumed to be unity).

\begin{figure}
\begin{center}
\includegraphics[height=7cm,keepaspectratio=true]{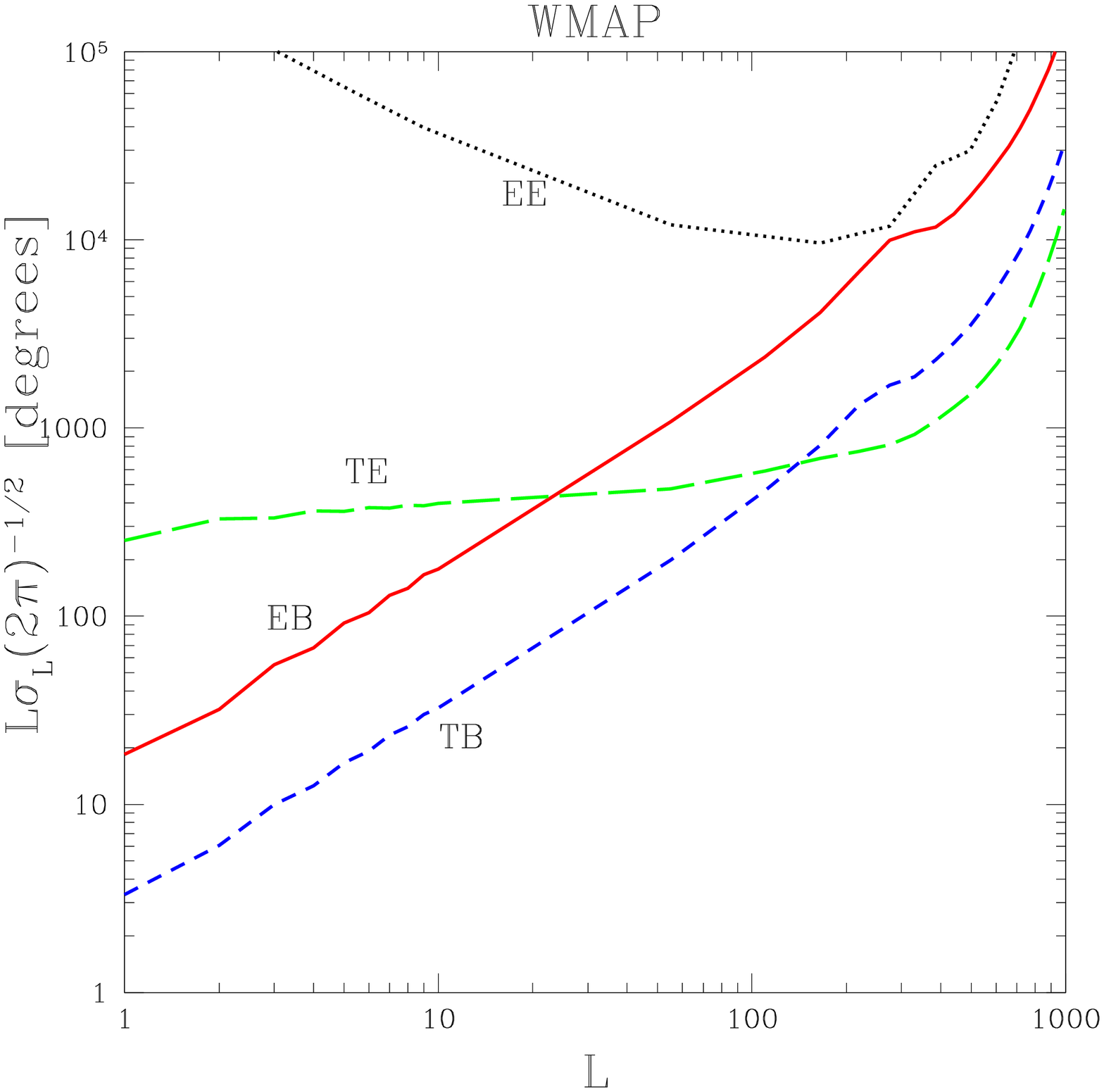}
\includegraphics[height=7cm,keepaspectratio=true]{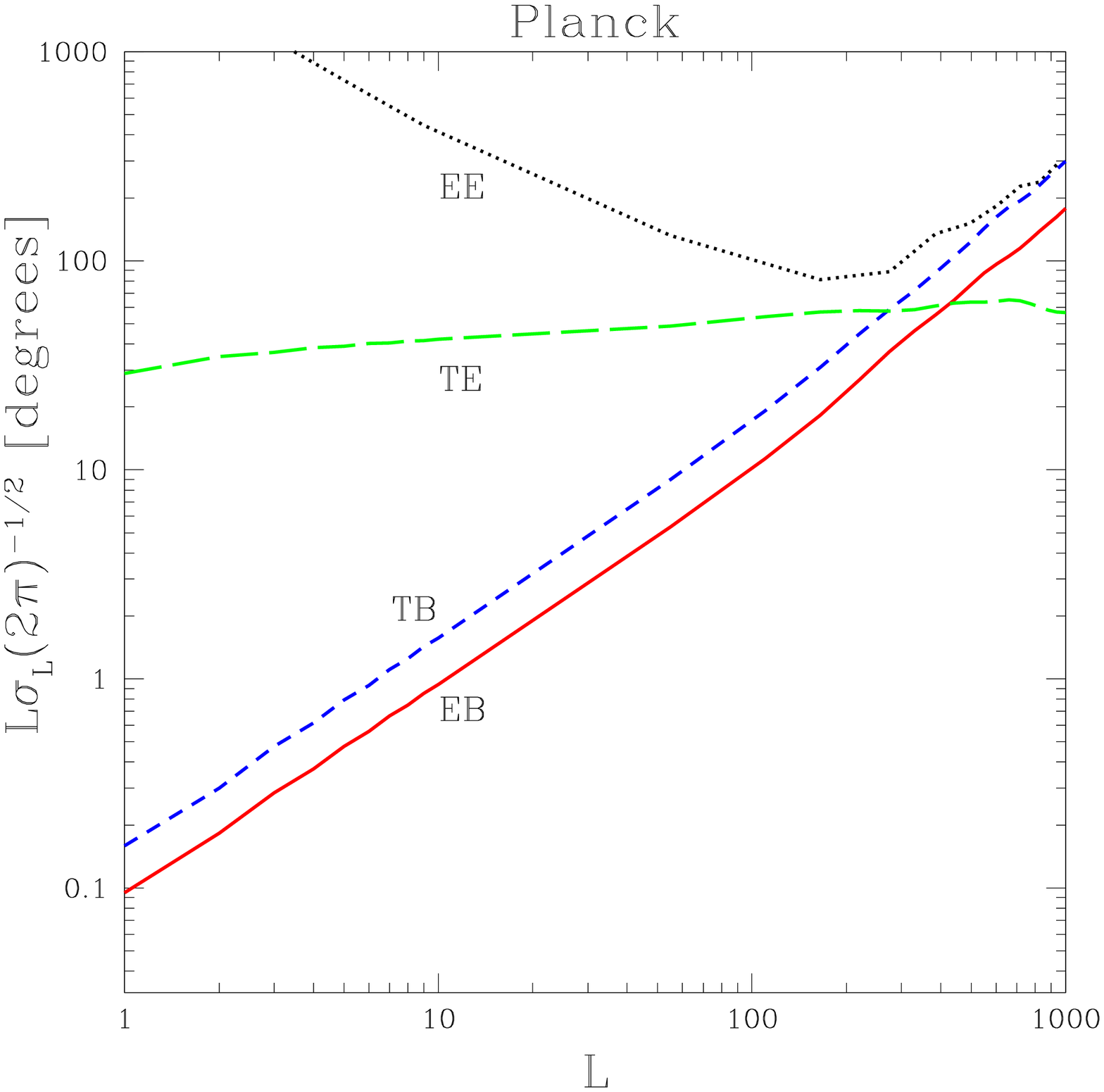}
\includegraphics[height=7cm,keepaspectratio=true]{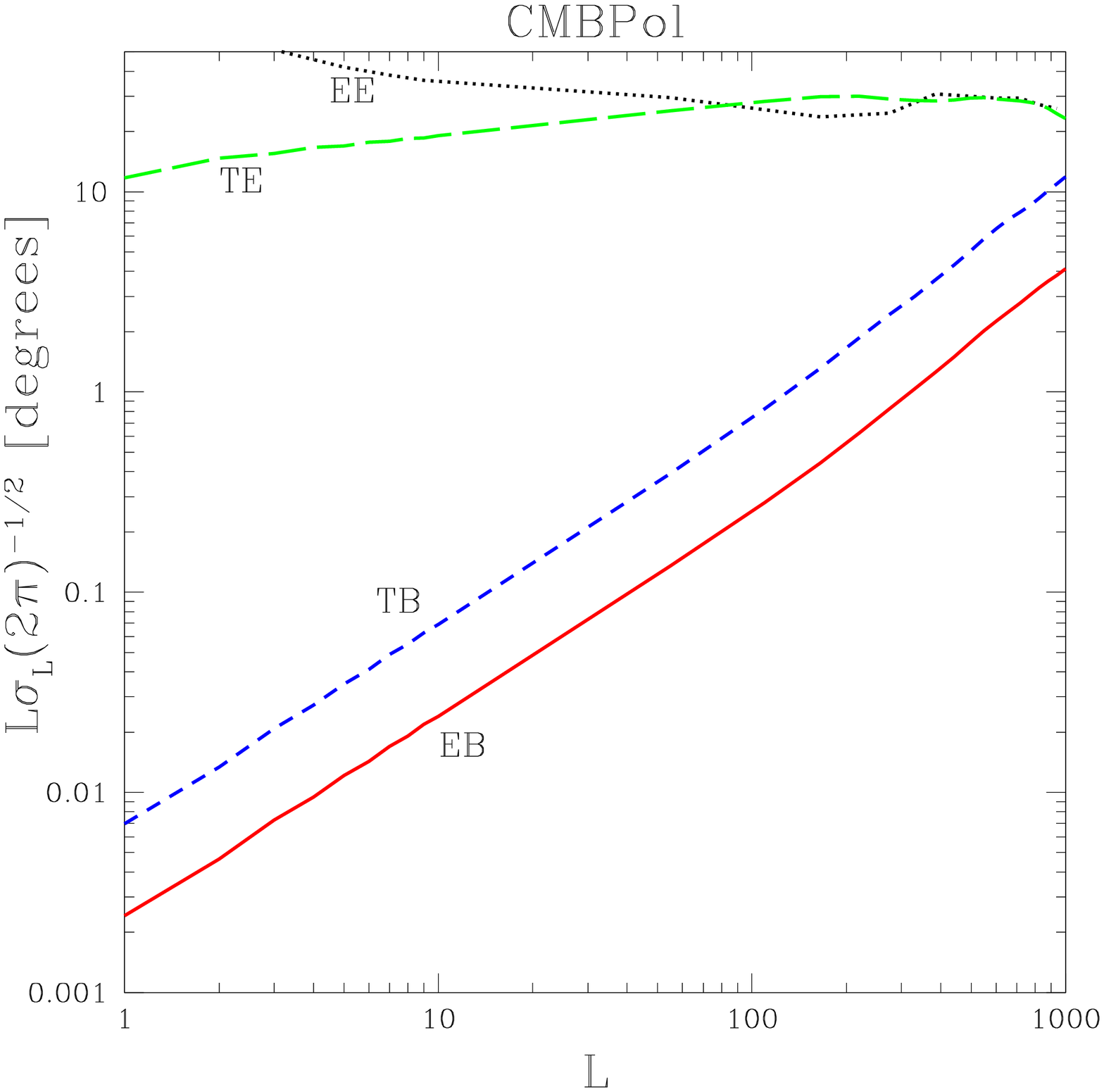}
\caption{Variances for the four rotation estimators for WMAP, Planck,
     and CMBPol are shown. We see that, at low multipoles, the
     lowest-error estimator for Planck and CMBPol comes from the
     EB correlation, and for WMAP from the TB
     correlation. \label{4estim}}%
\end{center}
\end{figure}

We evaluate the expressions for variances derived in the
previous Section, for each of the three experiments,
and show in Fig.~\ref{4estim} the numerical results for EB, EE,
TB, and TE correlations.

The first thing to notice about the variance levels in these
three experiments is that both the lower noise and the higher
resolution contribute to about two orders of magnitude
improvement in the sensitivity to a rotation
from WMAP to Planck and will lead to yet more than another
order-of-magnitude
improvement in CMBPol. This is illustrated in more
detail in Fig.~\ref{comp1000}, where the variances in the EB and
TB estimators are compared for all three
experiments.

\begin{figure}%
\begin{minipage}{10cm}%
\includegraphics[height=7cm,keepaspectratio=true]{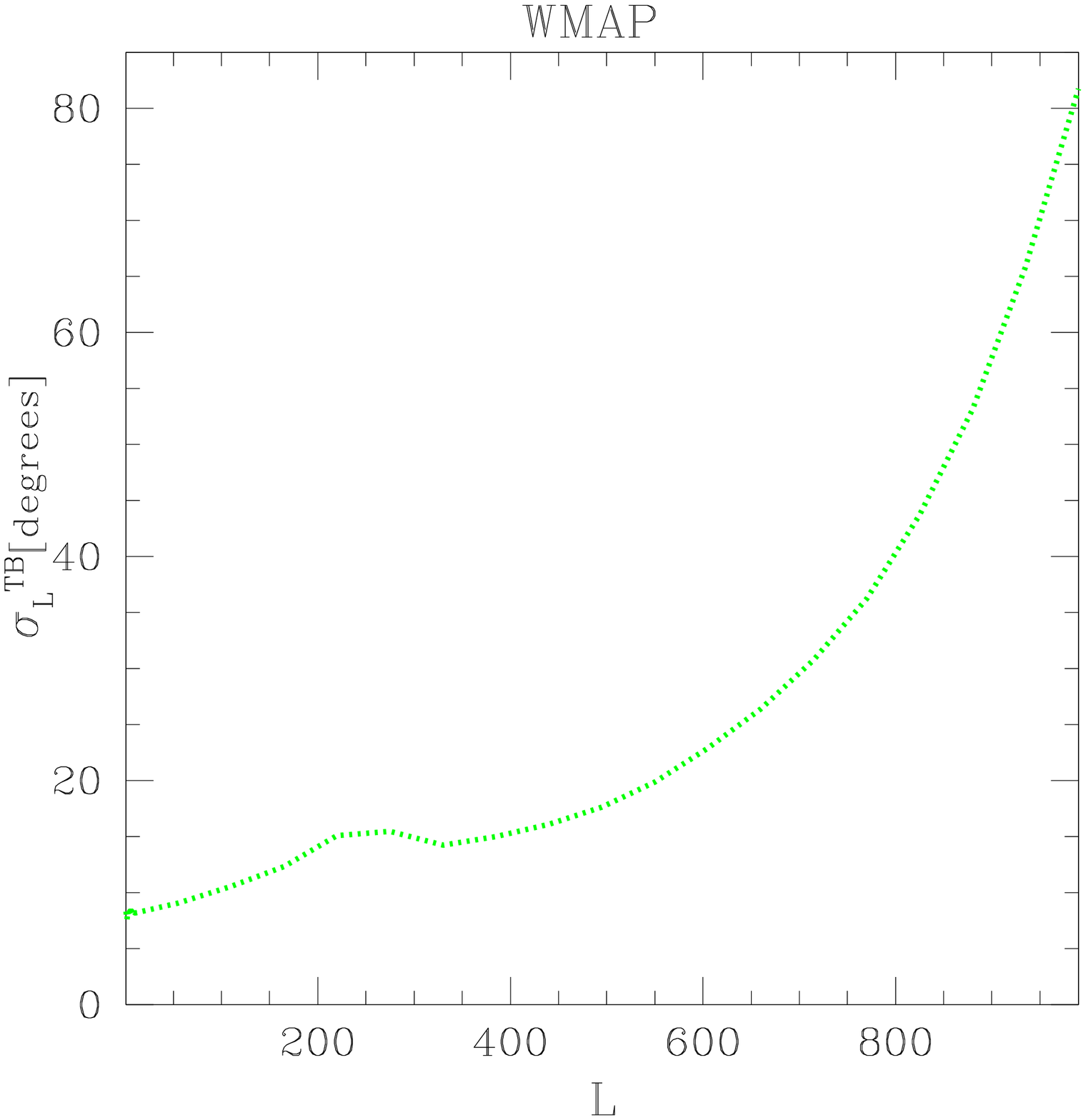}
\includegraphics[height=7cm,keepaspectratio=true]{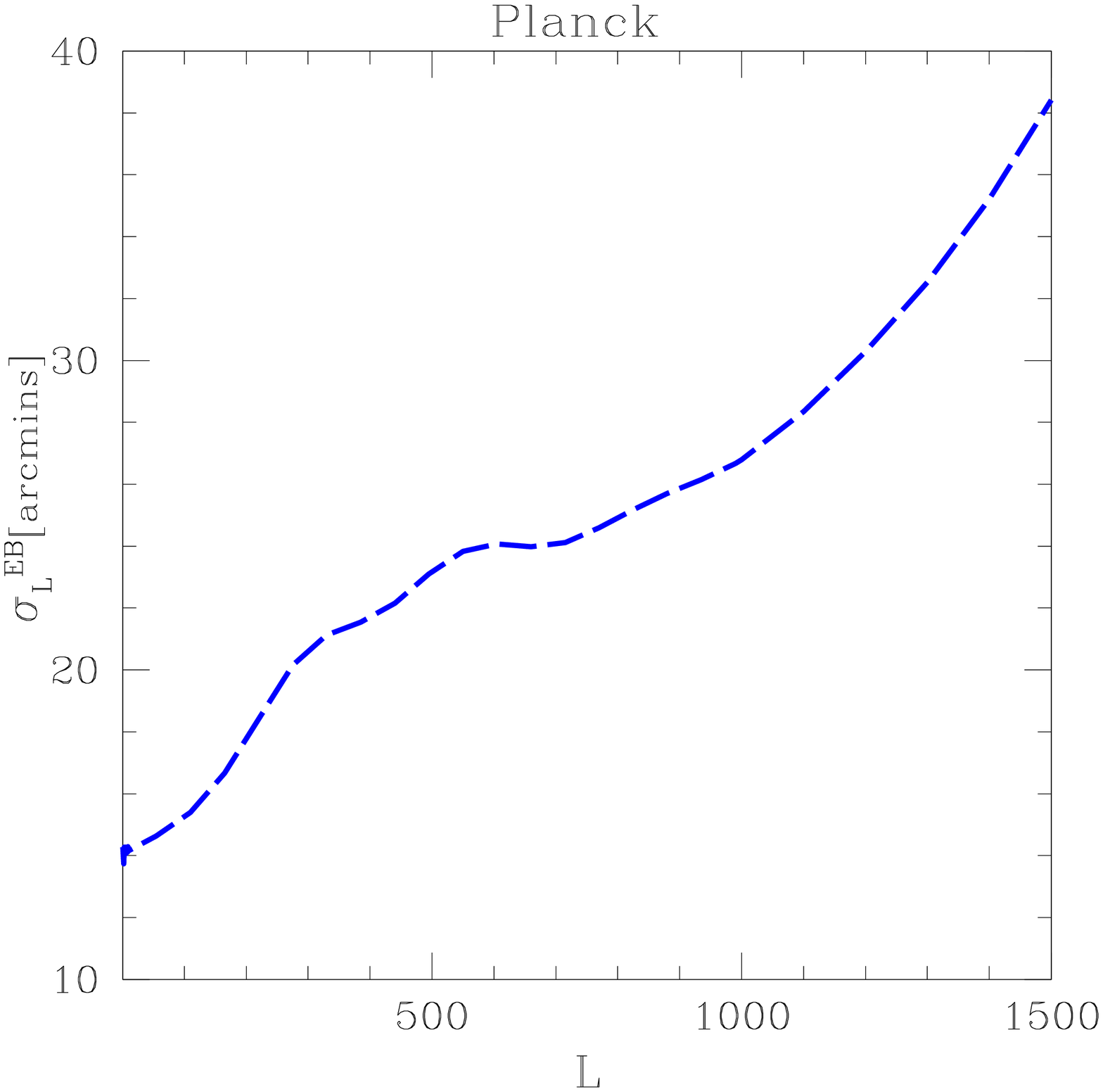}
\includegraphics[height=7cm,keepaspectratio=true]{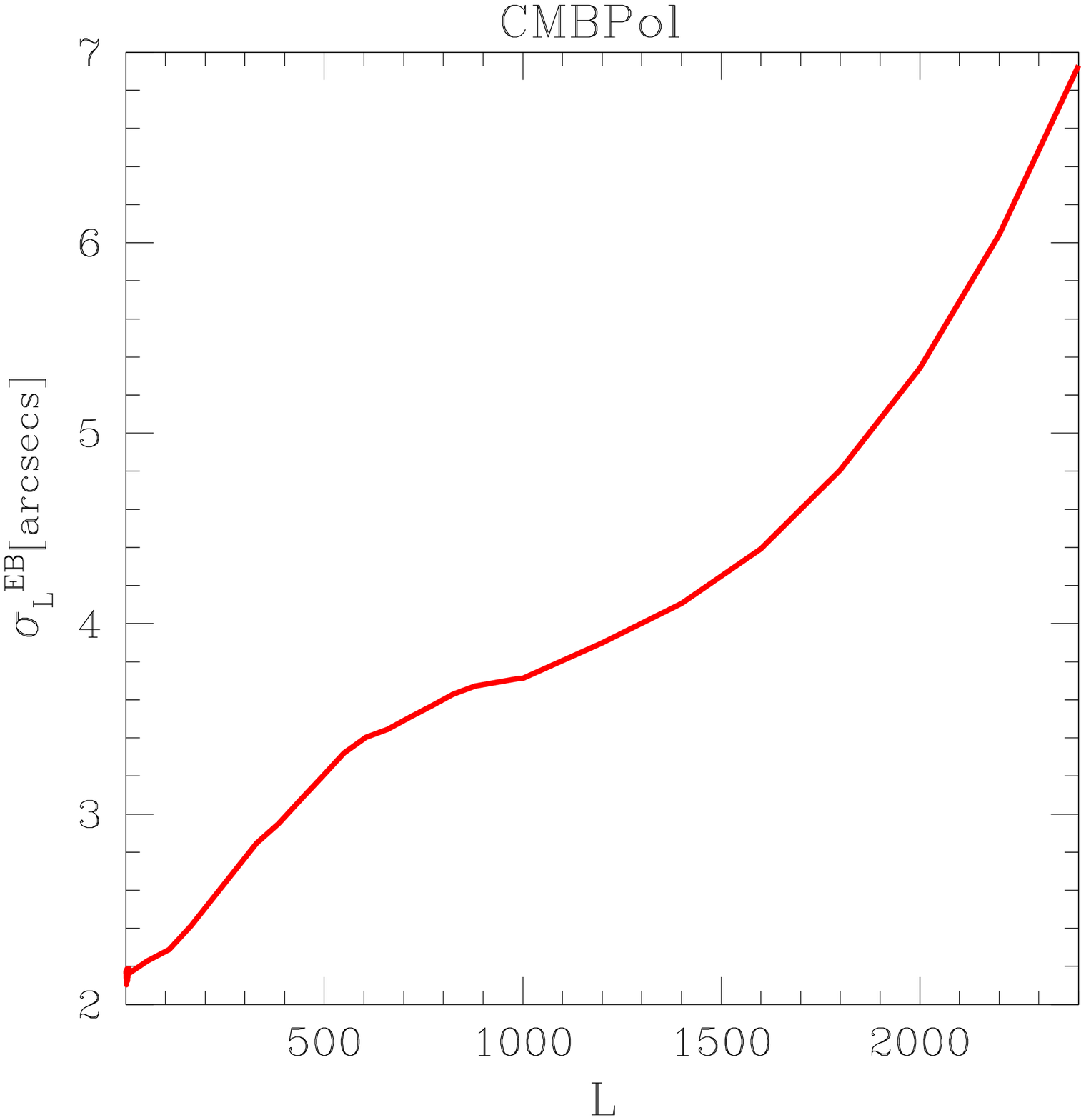}
\end{minipage}
\caption{The variances of the estimators for rotation are shown
     in linear scale, for WMAP (variance from the TB correlator),
     Planck (from EB) and CMBPol (from EB). The range of the plots is
     chosen so that it covers the resolution domain of the
     instrument. The exponential rise at high $L$, due to the window
     function, is visible. Note the difference in scale on the $y$-axis
     in all three panels.\label{lin}}%
\end{figure}

\begin{figure}%
\centering
\includegraphics[height=7cm,keepaspectratio=true]{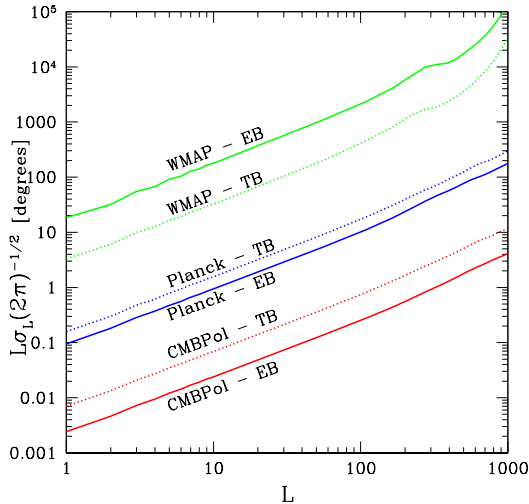}
\caption{The best constraints to rotation, i.e. the variances
     for estimators from the EB and TB correlations, are compared for
     three instruments. Planck gives about 2 orders of magnitude
     improvement in comparison to WMAP, and CMBPol is one order of
     magnitude better than Planck. 
\label{comp1000}}%
\end{figure}

From Fig.~\ref{4estim}, we see that, at low multipoles (below
$L$ of about $200$), the most sensitive estimators for all three
experiments will be those derived from the EB and TB
correlations. This comes about because of the absence of
any TB or EB correlation under the assumption (justified
largely by upper limits to the B mode amplitude) of no B modes
at the surface of last scatter.  
We separately look at the TB-estimator variance for
WMAP and the EB-estimator variances for Planck and CMBPol, in
Fig.~\ref{lin}. 

For WMAP, we find a TB-estimator variance of
$8.3^{\circ}$ at $L = 0$, which is consistent with the current
constraints on a uniform rotation\footnote{Note that we
calculate the multipole moments of rotation, so a uniform
rotation of angle $\alpha$ has $\alpha _{00}=\sqrt{4\pi}\alpha$. 
Thus, $\alpha_{00}=8.3^{\circ}$ is equivalent to $\alpha
=2.3^{\circ}$.} \cite{Feng:2006dp,Komatsu}
(see Appendix C). The dipole and quadrupole components of the rotation
have the same variances as the monopole ($L=0$), since the
variance remains fairly flat out to $L$ of about $100$, in all
three instruments.
Above $L \simeq 400$, the variance increases rapidly. This
happens when the exponential part of the window functions
dominates (due to the
finite angular resolution of the instrument). Also, since the
correlation angle for polarization is about $10$ times smaller
than that for temperature, the exponential tail in the EE-estimator
case becomes prominent at higher multipoles than in the TE
case.  The variance from the EB-estimator at $L=0$ is
$46.2^{\circ}$ for WMAP,
and thus not constraining. The TE-estimator variance quickly
drops below the EB-estimator variance (at $L\simeq20$) and
below the TB-estimator variance (at $L\simeq150$).  However, the
variance $\sigma_L$ at these $L$ is so large ($\gtrsim100$) that
the measurements are not at all constraining.  Similar
features are apparent in plots for the other two experiments.

For Planck, the variances of the EB and TB estimators are more
comparable, and the constraints to all rotation multipoles in
the range from $L$ of $0$ to about $300$ come from the EB
variance. At $L=0$ the variance is $14'$ and $24'$, for the EB-
and TB-estimator variances, respectively. Planck can thus
provide an order-of-magnitude better sensitivity to the
uniform rotation than the current WMAP sensitivity. For
high multipoles, above $L\simeq400$ or so, the TE-estimator variance
becomes the smallest one. At $L\simeq800$, a rapid rise in
all four variances is visible, due to the limitations in angular
resolution and the correlation angle of the polarization. 

For CMBPol, the EB-estimator variance is the smallest in the whole range
of multipoles from $0$ to $1000$. At $L=0$, the EB and TB values
are, respectively, $2.2"$ and $6.3"$, which is better than
Planck by more than an order of magnitude. Similarly to WMAP and
Planck variances, we observe a rapid rise in the variance above
$L\simeq1000$, corresponding to the resolution limitations
and/or the polarization correlation length. 

\begin{figure}%
\centering
\includegraphics[height=7cm,keepaspectratio=true]{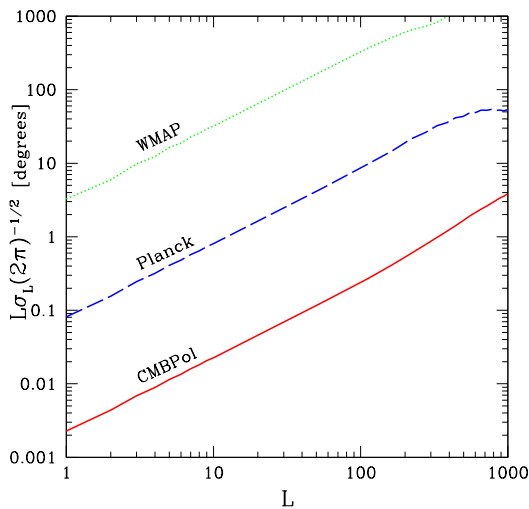}
\caption{The combined variance, for all four estimators for
     rotation, is shown for WMAP, Planck and CMBPol. Note that,
     due to the inverse-weight summing, the smallest of the four
     variances dominates the shape of the curves.\label{all}}%
\end{figure}

In Fig.~\ref{all} we show the variance for the combined
minimum-variance estimator, obtained from all four estimators,
with inverse-variance weighting.  We have not included the
covariance between the four estimators in this numerical
calculation.  However, this omission should make negligible
difference for the WMAP and CMBPol curves, where the variance is
determined primarily by TB and EB, respectively; the Planck
curve may be increased, but only slightly.  The run of this combined
variance with $L$ differs very little from the
smallest/constraining variance (that is TB in the case of WMAP
and EB in the case of Planck and CMBPol), because that term
dominates the sum.

\begin{figure}%
\centering
\includegraphics[height=7cm,keepaspectratio=true]{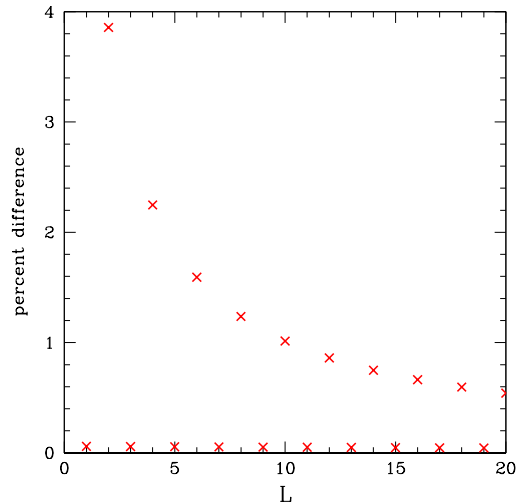}
\caption{Comparison between the full- and flat-sky expressions
     for the variance in the low-$L$ regime.  The discrepancy
     between the two is larger for even $L$ because of the
     additional contributions to the estimators from $l=l'$.}
\label{fig:compare}%
\end{figure}

As a check, we repeated our calculations for the EB-correlator using the flat-sky formulas from \S IV, and compared those results to the full-sky. The two
sets of variances are in good
agreement at high multipoles, where the flat sky makes a valid
approximation (better than a fraction of a percent for $L\gtrsim 50$). The
discrepancy increases up to $\sim4\%$ at lower $L$.
Fig.~\ref{fig:compare} compares the variances obtained from the
full- and flat-sky treatments at low $L$. Finally, we re-did the full-sky calculations using the same instrumental parameters as in Ref.~\cite{Yadav:2009eb}; these are also in good agreement with our results at high L's (better than $7\%$ for $L\gtrsim 50$, and about $11\%$ at $L=0$).
\section{Conclusions and Summary}

Exotic mechanisms, such as quintessence fields that couple to
the pseudoscalar of electromagnetism, could rotate the linear
polarization of the CMB. In this paper, we derive the complete
set of minimum-variance estimators for a
position-dependent rotation of the CMB polarization, and thus
provide a recipe, given a full-sky map, for measuring the
rotation angle as a function of position on the sky.  We also
evaluate the variances with which the rotation-angle
spherical-harmonic coefficients can be measured for WMAP, Planck, and CMBPol.  Our results
indicate that EB and TB correlations will provide more sensitive
probes of the rotation angle than TE and EE correlations, and
that EB becomes better, relative to TB, as the instrumental
noise is reduced.  We have checked that our results for the
variances recover prior results, both analytically and
numerically, in the flat-sky limit.  As an additional check,
Appendix C shows that our expressions for the rotation-angle
variance recovers that expected for a uniform rotation.
Appendix B shows that parity considerations can be used to
distinguish the effects of rotation and of weak lensing on the
CMB temperature/polarization map.
 
Now that we have elucidated the all-sky formalism, the next step
will be to work out algorithms for recovery of the
rotation angle for a map with partial sky coverage.  We
anticipate that analogous techniques for determining the
cosmic-shear deflection angle may be adapted for this purpose.

\begin{acknowledgments}
VG thanks Itamar Yaakov and Michael Kesden for useful comments. This work was
supported by DoE DE-FG03-92-ER40701 and the Gordon and Betty
Moore Foundation.
\end{acknowledgments}


\section*{APPENDIX A: Some Useful Formulas}

Some formulas used in \S II and \S III are included here.
The metric tensor and its inverse on a unit 2-sphere (i.e. on
the \textit{sky}) is
\begin{equation}
g = \left( {\begin{array}{*{20}c}
   1 & 0  \\
   0 & {\sin ^2 \theta }  \\
 \end{array} } \right),
\end{equation}
\begin{equation}
g^{ - 1}  = \left( {\begin{array}{*{20}c}
   1 & 0  \\
   0 & {\frac{1}
{{\sin ^2 \theta }}}  \\
 \end{array} } \right).
\end{equation}
The Levi-Civita tensor on a unit 2-sphere is
\begin{equation}
\varepsilon  = \left( {\begin{array}{*{20}c}
   0 & {\sin \theta}  \\
   { - \sin \theta } & 0  \\
 \end{array} } \right).
\end{equation}
An orthonormal basis on a unit 2-sphere is
\begin{equation}
\hat e_\theta   = \left( \begin{gathered}
  1 \hfill \\
  0 \hfill \\ 
\end{gathered}  \right), \qquad \hat e_\phi   = \left( \begin{gathered}
  0 \hfill \\
  \sin \theta  \hfill \\ 
\end{gathered}  \right).
\end{equation}
Some useful properties of the Wigner 3j symbols and related quantities include
\begin{equation}
\left( {\begin{array}{*{20}c}
   {l_1 } & {l_2 } & {l_3 }  \\
   {m_1 } & {m_2 } & {m_3 }  \\
 \end{array} } \right) = ( - 1)^{l_1  + l_2  + l_3 } \left( {\begin{array}{*{20}c}
   {l_1 } & {l_2 } & {l_3 }  \\
   { - m_1 } & { - m_2 } & { - m_3 }  \\
 \end{array} } \right),
\end{equation}
\begin{equation}
\left( {\begin{array}{*{20}{c}}
   {{l_1}} & {{l_2}} & {{l_3}}  \\
   {{m_1}} & {{m_2}} & {{m_3}}  \\
 \end{array} } \right) = \left( {\begin{array}{*{20}{c}}
   {{l_2}} & {{l_3}} & {{l_1}}  \\
   {{m_2}} & {{m_3}} & {{m_1}}  \\
 \end{array} } \right),
\end{equation}
\begin{equation}
\left( {\begin{array}{*{20}{c}}
   {{l_1}} & {{l_2}} & {{l_3}}  \\
   {{m_1}} & {{m_2}} & {{m_3}}  \\
 \end{array} } \right) = {( - 1)^{{l_1} + {l_2} + {l_3}}}\left( {\begin{array}{*{20}{c}}
   {{l_2}} & {{l_1}} & {{l_3}}  \\
   {{m_2}} & {{m_1}} & {{m_3}}  \\
 \end{array} } \right),
\end{equation}
\begin{equation}\left( {\begin{array}{*{20}{c}}
   l & l & 0  \\
   m & { - m} & 0  \\
 \end{array} } \right) = \frac{{{{( - 1)}^{l - m}}}}
{{\sqrt {2l + 1} }},
\end{equation}
\begin{equation}
m_1+m_2+m_3 \ne 0 \Rightarrow \left( {\begin{array}{*{20}c}
   l_1 & l_2 & {l_3 }  \\
   m_1 & m_2 & m_3  \\
 \end{array} } \right) = 0,
\end{equation}
\begin{equation}
\begin{gathered}
 \sum\limits_{{m_1}{m_2}} {} (2{l_3} + 1)\left( {\begin{array}{*{20}{c}}
   {{l_1}} & {{l_2}} & {{l_3}}  \\
   {{m_1}} & {{m_2}} & {{m_3}}  \\
 \end{array} } \right)\left( \begin{array}{*{20}{c}}
   {{l_1}} & {{l_2}} & {{l_3}'}  \\
   {{m_1}} & {{m_2}} & {{m_3}'}  \\
 \end{array}  \right) \hfill \\
= {\delta_{{l_3}{l_3}'}}{\delta_{{m_3}{m_3}'}}.
\end{gathered} 
\end{equation}
The relation between spherical harmonics and Wigner 3j symbols is
\begin{equation}
\begin{gathered}
\int {d\hat
n{Y_{{l_1}{m_1}}(\hatn)}{Y_{{l_2}{m_2}}(\hatn)}{Y_{{l_3}{m_3}}(\hatn)}} \\
  = \sqrt {\frac{{(2{l_1} + 1)(2{l_2} + 1)(2{l_3} + 1)}} 
{{4\pi }}} \\ \times \left( {\begin{array}{*{20}{c}}
   {{l_1}} & {{l_2}} & {{l_3}}  \\
   {{m_1}} & {{m_2}} & {{m_3}}  \\
 \end{array} } \right)\left( {\begin{array}{*{20}{c}}
   {{l_1}} & {{l_2}} & {{l_3}}  \\
   0 & 0 & 0  \\
 \end{array} } \right)
\end{gathered}
\end{equation}


\section*{APPENDIX B: Rotation vs. Weak Lensing}

The effects of rotation and weak lensing
\cite{Hu:2000ee,lensing} on polarization are
orthogonal and can thus be distinguished geometrically with a
full-sky map.  For example, if we start off with a pure E mode at the surface of
last scatter, rotation induces a B mode, given by
Eq.~(\ref{InducedBMode_FromPureE_General}), where the only
non-zero terms are those that satisfy
$L+l+l_2=$even. However, if we analyse the effect of weak
lensing (see Ref.~\cite{lensing}), a pure E mode
polarization tensor changes by  
\begin{equation}
\delta {P_{ab}} = ({\nabla _c}\varphi )({\nabla ^c}{P_{ab}}),
\label{PolarizationChangeByLensing}
\end{equation}
where $\varphi$ is the projection of the gravitational potential along the line of sight. Thus, the B mode induced by weak lensing is
\begin{equation}
\begin{gathered}
\delta B_{lm}  = \int {d\hat n\delta P_{ab} (\hat n)Y^{B*,ab} _{(lm)} } (\hat n) \hfill \\  
= 2\sum\limits_{LM} {\sum\limits_{l_2 m_2 } }\varphi _{LM} E_{l_2 m_2 }\hfill \\
 \int {d\hat n}  Y^{B*,ab} _{(lm)} (\nabla _c Y_{(LM)}) (\nabla ^c Y^E _{(l_2 m_2 )ab}) .
\end{gathered}
\label{InducedBMode_FromPureE_Lensing}
\end{equation}
The parity of the spherical harmonic $Y_{LM}$ is $(-1)^L$. The
parity of the E-mode term in the integral is $(-1)^{l_2}$, and the parity of the B-mode term is $(-1)^{l+1}$. The parity of the integrand in
Eq.~(\ref{InducedBMode_FromPureE_Lensing}) is then
$(-1)^{L+l+l_2+1}$. Therefore, the integral is non-vanishing
only for terms that satisfy $L+l+l_2=$odd. 

We conclude that the rotation induces B modes that satisfy
$L+l+l_2=$even, while for weak lensing we have $L+l+l_2=$odd
[see Eqs.~(\ref{InducedBMode_FromPureE_General}) and
(\ref{InducedBMode_FromPureE})]. Thus, the two effects can be
entirely separated.

Strictly speaking, this orthogonality between lensing and
rotation occurs only at linear order in the rotation and lensing
amplitudes in the limit that lensing and rotation are both small.
If a lensed field is then rotated, and/or if a rotated field is
then lensed, then the orthogonality will break down.  However,
this will occur only with an amplitude that is proportional to
the product of the lensing and rotation amplitudes.  We have
here implicitly assumed this to be small and leave the full
treatment of this higher-order effect for future work.


\section*{APPENDIX C: Uniform Rotation Cross-Check}

We can perform a cross-check of our formulas for variances of
the rotation, by analyzing only the $L=0$ term, where all the
representation-theory coefficients can be readily evaluated.  We
do so for the EB case.
From Eqs.~(\ref{sigma_XGH_in_EB}) and (\ref{XDefinition_inEB})
(for $l=l'$), after evaluating the coefficients for $L=0$, we
get the variance of the uniform-rotation estimator to be
\begin{equation}
{\begin{gathered}
  {(\sigma _{00}^{})^{-2}} = \frac{1}{\pi} \sum\limits_l
  {\frac{{{[C_l^{EE}{{({W_l})}^2}]}^2 (2l+1)}}{C
  _{l}^{EE\text{,map}}C _{l}^{BB\text{,map}}}}  \hfill \\ 
\end{gathered}} .
\label{sigma_00}
\end{equation}

We can see that this is indeed the right expression for uniform
rotation, by noting that the B mode induced by small rotation,
from a pure E mode, is given by
Eq.~(\ref{PolarizationChangeBySmallRotation}). The induced EB
power spectrum in that case is
\begin{equation}
     C_l^{EB} = 2{\alpha}C_l^{EE},
\label{CEB_uniformRotation}
\end{equation}
and estimators for the rotation can be expressed for each $lm$ pair as
\begin{equation}
     {{\hat \alpha }_{}} =
     \frac{{E_{lm}^{\text{map}}B_{lm}^{\text{map}}}}
     {{2C_l^{EE}W_l^2}}.
\label{alpha_uniformRotation}
\end{equation}
The variance of rotation is then calculated from all $lm$ pairs as
\begin{equation}
\begin{gathered}
     \sigma _{}^{ - 2} = \sum\limits_{l = 0}^\infty
     {\sum\limits_{m =  - l}^l {} } \frac{1} 
     {{\left\langle {{{({{\hat \alpha }_{}})}^2}} \right\rangle }}
     = \sum\limits_{l} (2l +
     1)\frac{{4{{[C_l^{EE}{{({W_l})}^2}]}^2}}}
     {{C_l^{EE{\text{,map}}}C_l^{BB{\text{,map}}}}},
\end{gathered}
\label{variance_uniformRotation}
\end{equation}
where the factor of $(2l+1)$ comes from the sum over $m$,
because the terms in the sum have effectively only index $l$. For $L=0$,
Eq.~(\ref{variance_uniformRotation}) reduces to
Eq.~(\ref{sigma_00}), once the factor of $\sqrt{4\pi}$ by which
$\alpha$ and $\alpha _{00}$ differ is taken into account.
\section*{APPENDIX D: Contributions of Multipole pairs to
Rotation Estimator}
\begin{figure}%
\centering
\begin{minipage}{10cm}%
\includegraphics[height=5.5cm,keepaspectratio=true]{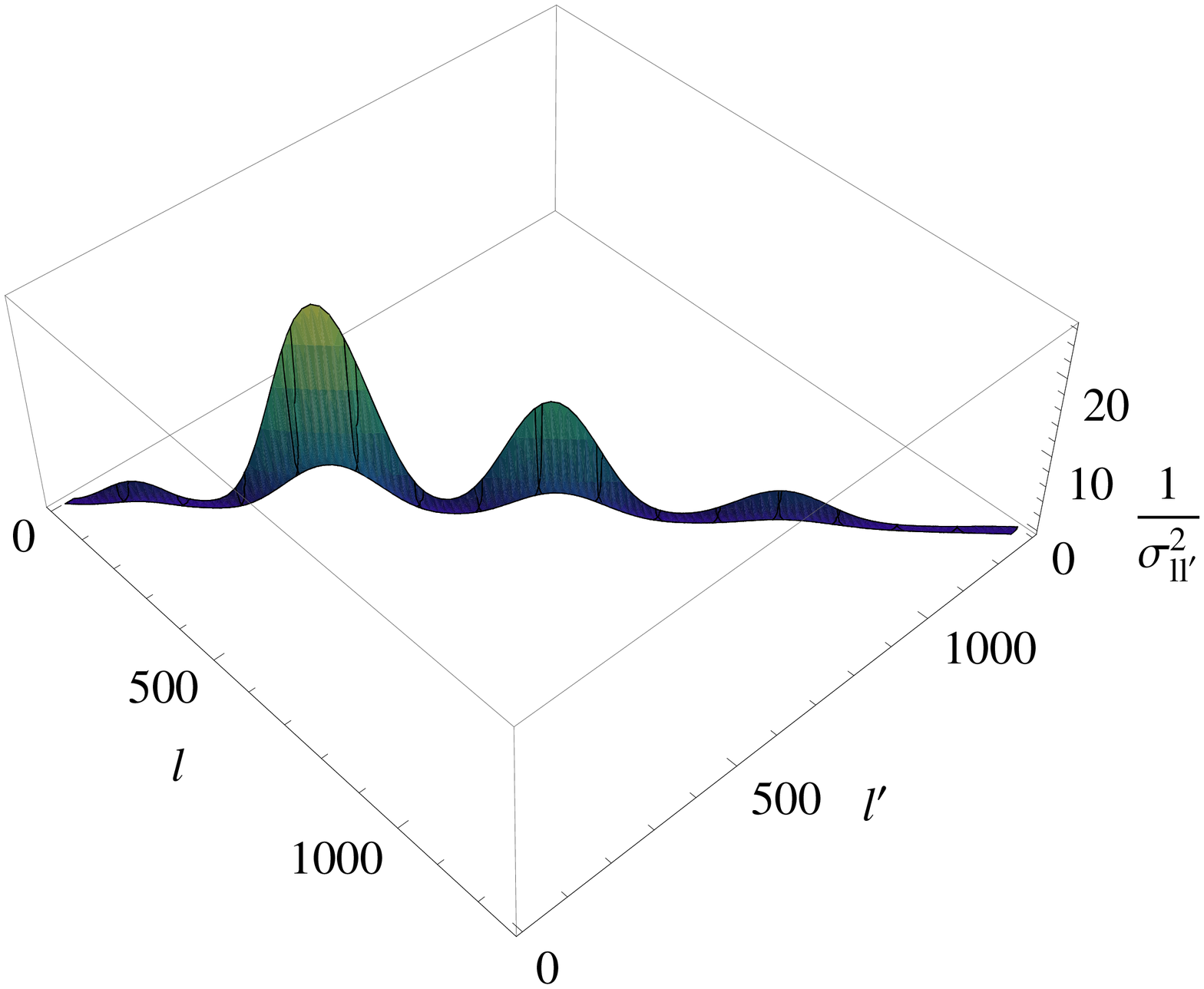}
\includegraphics[height=5.5cm,keepaspectratio=true]{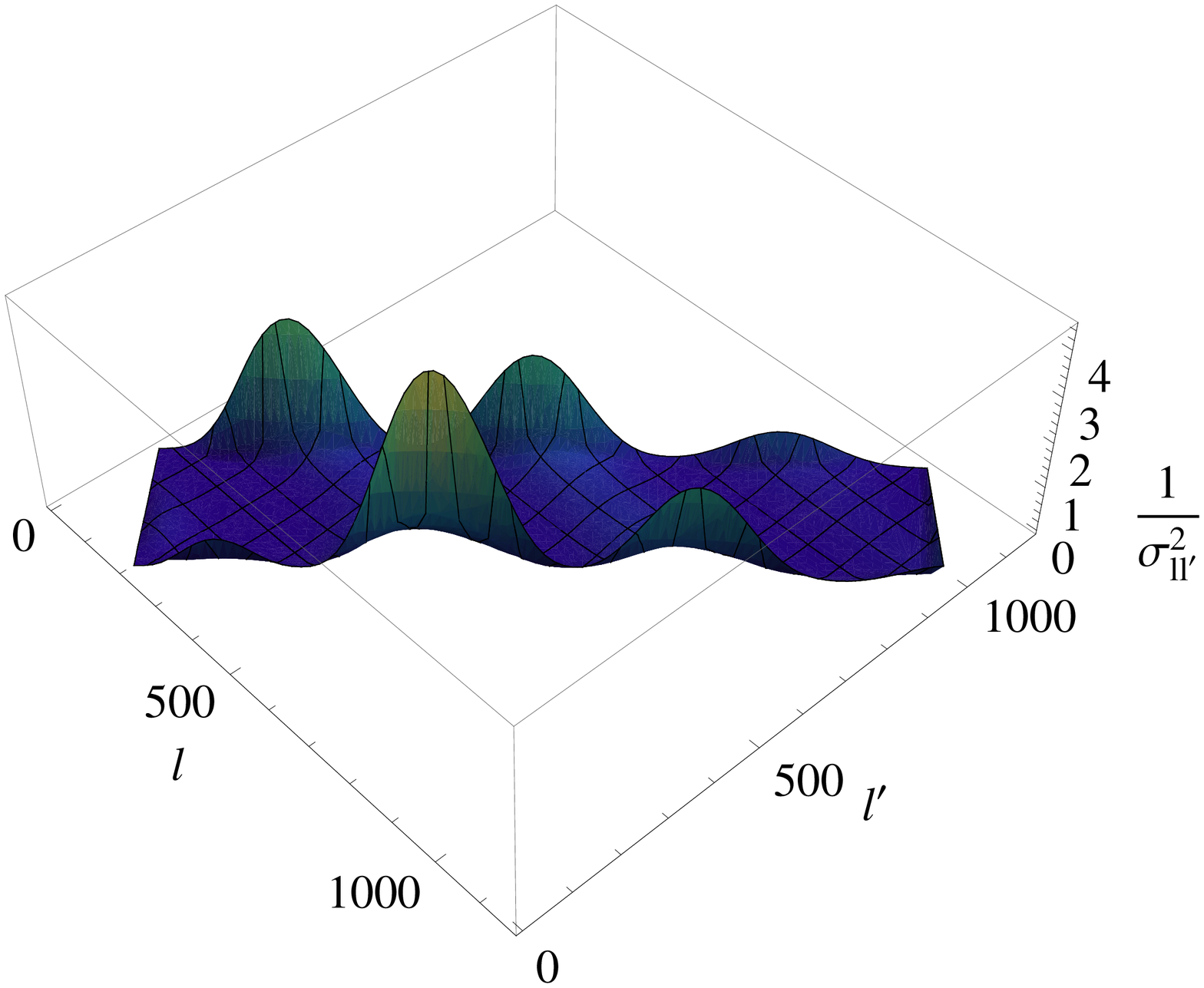}
\includegraphics[height=5.5cm,keepaspectratio=true]{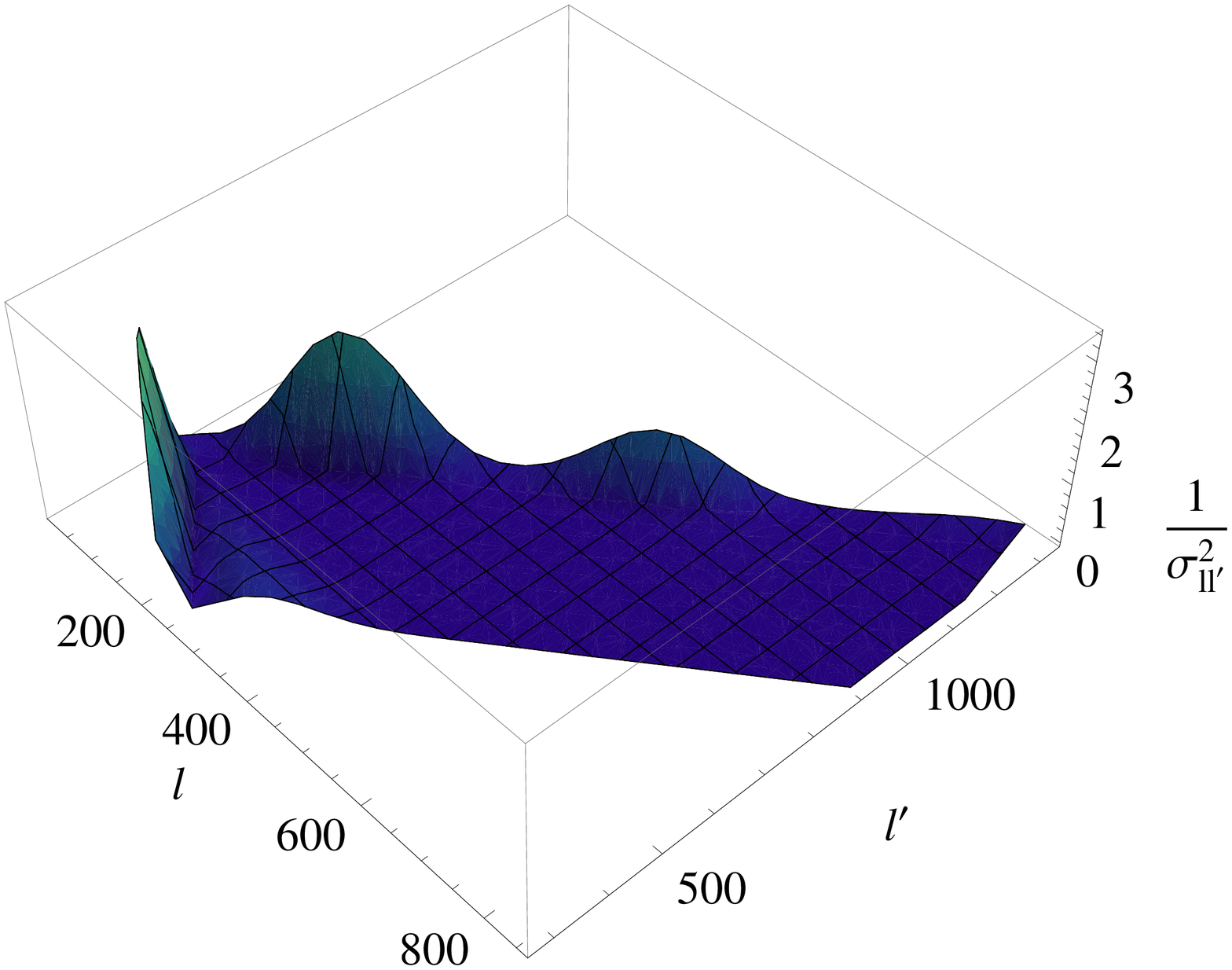}
\end{minipage}
\caption{The summands in Eq.~(\ref{BEnoise}) are plotted versus
     $l$ and $l'$. Plots for three different rotation multipole
     coefficients are shown: $L=20$ (top panel), $L=200$ (middle
     panel), and $L=500$ (bottom panel). The plots suggest that
     the major contribution to any given rotation multipole
     coefficient comes from the $ll'$ estimators that correspond
     to the strongest peaks in the EE power spectrum above the
     resolution limit of the instrument.\label{contributions}}%
\end{figure}
\begin{figure}
\includegraphics[height=7cm,keepaspectratio=true]{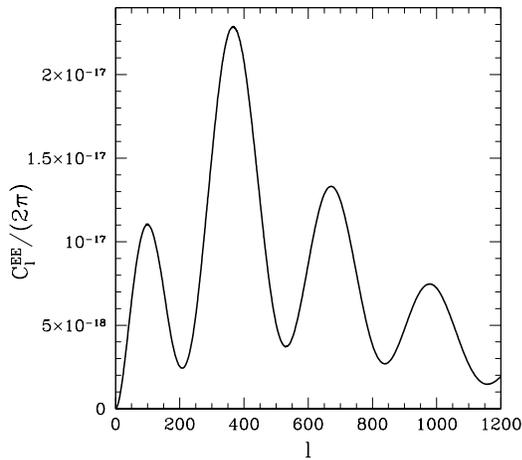}
\caption{The EE power spectrum used for numerical estimates in
     \S V is shown. The main peaks of the power spectrum are
     visible at $L\sim 150$, $400$, $700$, and
     $700$.\label{power_EE}}
\end{figure}

In Fig.~\ref{contributions}, we show the summand in
Eq.~(\ref{BEnoise}), versus $l$ and $l'$, for $L=20$, $200$, and
$500$. These terms are the weights with which each $ll'$
pair contributes to the EB estimator $\hat{\alpha}_{LM}$ for
Planck (see Eq.~(\ref{BEestimator})). Fig.~\ref{contributions}
therefore illustrates which multipoles in the EE power spectrum
are expected to contribute most to the three chosen rotation
multipoles. From the Figure, we see that the region allowed by
triangle inequalities grows with $L$, and that the peaks along
$l'$ correspond to the peaks in the EE power spectrum. There is
also an overall exponential decay from the window
functions, with a characteristic scale of $l'\sim 1000$, beyond
which the distributions fall to zero.

Overall, the EE multipoles that affect the estimate of the
rotation at $L=20$ and $200$ seem to be predominantly those
that correspond to the strongest peaks in the EE power spectrum,
below the resolution limit of $l'\sim 1000$ (i.e. $l'$ of about
$400$, $700$, and $1000$; see Fig.~\ref{power_EE}). The relative
contributions are shown in the top two panels of
Fig.~\ref{contributions}. The bottom panel of the same Figure
shows the $L=500$ case. There, we see that the largest
contribution is local, i.e. it comes from non-flat triangles
at multipoles below $l'\sim 500$, and the secondary contribution
comes from the EE power-spectrum peaks at $l'$ of about $700$
and $1000$ (see Fig.~\ref{power_EE}).

\end{document}